\documentclass[lettersize,journal]{IEEEtran}
\usepackage[table,xcdraw]{xcolor}
\usepackage{amsmath,amsfonts,amssymb}
\usepackage{algorithmic}
\usepackage[ruled,vlined,linesnumbered,lined]{algorithm2e}

\SetCommentSty{mycommfont}
\usepackage{array}
\usepackage[caption=false,font=footnotesize,
labelfont=rm,textfont=rm]{subfig}
\usepackage{textcomp}
\usepackage{stfloats}
\usepackage{url}
\usepackage{verbatim}
\usepackage{graphicx}
\usepackage{cite}
\usepackage{graphicx}\graphicspath{{../figure/}{figure_case_study/}}
\usepackage{tikz}
\usepackage{siunitx}
\usepackage{cases}
\usepackage{newtxmath}
\usepackage{balance}
\usepackage{threeparttable}
\usepackage{cuted}
\usepackage{multirow}
\newcommand{\bm}{\dot{\boldsymbol{m}}}
\newcommand{\bmq}{\dot{\boldsymbol{m}}^\text{in}}

\newcommand{\mq}{\dot{m}^\text{in}}

\newcommand{\bM}{\dot{\boldsymbol{M}}}

\newcommand{\bMq}{\dot{\boldsymbol{M}}^\text{in}}
\newcommand{\m}{\dot{m}}
\newcommand{\M}{\dot{M}}
\newcommand{\Mq}{\dot{M}^{\text{in}}}

\newcommand{\bMqast}{\dot{\boldsymbol{M}}^{\text{in}\ast}}

\newcommand{\bT}{\boldsymbol{T}}
\newcommand{\btau}{\boldsymbol{\tau}}
\newcommand{\bphi}{\boldsymbol{\phi}}
\newcommand{\bPhi}{\boldsymbol{\Phi}}

\newcommand{\bp}{\boldsymbol{p}}
\newcommand{\bP}{\boldsymbol{P}}

\newcommand{\bq}{\boldsymbol{q}}
\newcommand{\bQ}{\boldsymbol{Q}}

\newcommand{\bee}{\boldsymbol{e}}
\newcommand{\bff}{\boldsymbol{f}}

\newcommand{\bw}{\boldsymbol{w}}
\newcommand{\bW}{\boldsymbol{W}}

\newcommand{\bE}{\boldsymbol{E}}
\newcommand{\bF}{\boldsymbol{F}}

\newcommand{\bU}{\boldsymbol{U}}

\newcommand{\md}{\mathrm{diag}}

\newcommand{\bTs}{\bT^\text{s}}
\newcommand{\bTsast}{\bT^{\text{s}\ast}}
\newcommand{\bTr}{\bT^\text{r}}
\newcommand{\bTrast}{\bT^{\text{r}\ast}}

\newcommand{\Ts}{T^{\text{s}}}
\newcommand{\Tr}{T^{\text{r}}}

\newcommand{\taus}{\tau^\text{s}}
\newcommand{\taur}{\tau^\text{r}}
\newcommand{\tauouts}{\tau^\text{out,s}}
\newcommand{\tauoutr}{\tau^\text{out,r}}
\newcommand{\tauins}{\tau^\text{in,s}}
\newcommand{\tauinr}{\tau^\text{in,r}}
\newcommand{\tauoutsr}{\tau^\text{out,s/r}}
\newcommand{\tauinsr}{\tau^\text{in,s/r}}
\newcommand{\btauinsr}{\boldsymbol{\tau}^\text{in,s/r}}

\newcommand{\Touts}{T^{\text{out,s}}}
\newcommand{\Toutr}{T^{\text{out,r}}}
\newcommand{\Tins}{T^{\text{in,s}}}

\newcommand{\bTouts}{\bT^{\text{out,s}}}
\newcommand{\bToutr}{\bT^{\text{out,r}}}
\newcommand{\bTins}{\bT^{\text{in,s}}}
\newcommand{\bTinr}{\bT^{\text{in,r}}}

\newcommand{\bPhiast}{\bPhi^\ast}
\newcommand{\bx}{\boldsymbol{x}}

\newcommand{\bX}{\boldsymbol{X}}

\newcommand{\by}{\boldsymbol{y}}
\newcommand{\bY}{\boldsymbol{Y}}

\newcommand{\bz}{\boldsymbol{z}}
\newcommand{\bZ}{\boldsymbol{Z}}

\newcommand{\sumnol}{\sum\nolimits}

\usepackage[hidelinks]{hyperref}

\begin{document}
\title{Non-iterative Calculation of Quasi-Dynamic Energy Flow in the Heat and Electricity Integrated Energy Systems}
\author{Ruizhi~Yu,~\IEEEmembership{Student~Member,~IEEE, }Wei~Gu,~\IEEEmembership{Senior~Member,~IEEE, }Hai~Lu, Shuai~Yao,~\IEEEmembership{Student~Member,~IEEE, }Suhan~Zhang,~\IEEEmembership{Student~Member,~IEEE, }Shuai~Lu,~\IEEEmembership{Member,~IEEE, }Shixing~Ding,~\IEEEmembership{Student~Member,~IEEE, }Enbo~Luo
    \thanks{This work was supported by the National Key Research and Development Program of China (Grant No. 2020YFE0200400), and in part by the Natural Science Foundation of Jiangsu Province (Grant No. BK20200013).}
    \thanks{R.Z. Yu, W. Gu, S. Yao, S.H. Zhang and S. Lu are with the School of Electrical Engineering, Southeast University, Nanjing, Jiangsu, 210096 China. \emph{Corresponding author: Wei Gu}, e-mail: wgu@seu.edu.cn.}
    \thanks{H. Lu, E.B. Luo are with the Yunnan Power Grid Co., Ltd, Kunming, Yunnan 650217, China.}
    \thanks{S.X. Ding is with the School of Cyber Science and Engineering, Southeast University, Nanjing 210096, China.}
}
\maketitle
\begin{abstract}
Quasi-dynamic energy flow calculation is an indispensable tool for the heat and electricity integrated energy system (HE-IES) analysis. One solves the nonlinear partial differential algebraic equations to obtain thermal, hydraulic and electric variations. However, mainstream iteration solvers face the challenges of inefficiency and bad robustness. For one thing, the frequent update and factorization of Jacobian matrices utilize high CPU time. For another, the per-step iteration numbers grow exponentially as the system loading level creeps up. This paper presents a novel non-iterative algorithm for the quasi-dynamic energy flow calculation.
The kernel of the proposed algorithm is to transform these nonlinear equations into linear recursive ones, by solving which, we obtain explicit closed-form solutions of unknown variables. In each step, the proposed algorithm requires only one matrix factorization and fixed times of arithmetic operations regardless of the loading levels, so that it achieves small and consistent per-step time costs. A semi-discrete scheme is used in PDE solution to avoid dissipative and dispersive errors that are often overlooked in previous literature. To ensure convergence, we also propose to control the temporal step sizes adaptively by estimating the simulation errors. 
Case studies showed that the proposed method manifested efficient and robust time performance compared with the iterative algorithms, and meanwhile preserved high accuracy.
\end{abstract}
\section*{Nomenclature}
\addcontentsline{toc}{section}{Nomenclature}
    \subsection{Abbreviation}
\begin{IEEEdescription}[\IEEEusemathlabelsep\IEEEsetlabelwidth{HE-IES}]
    \item[DHS] District heating system.
    \item[DT] Differential transformation.
    \item[EPS] Electric power system.
    \item[FDM] Finite difference method.
    \item[HE-IES] Heat and electricity integrated energy system.
    \item[IU] Implicit upwind scheme.
    \item[NM] Node method.
    \item[ODE] Ordinary differential equation.
    \item[PDE] Partial differential equation.
    \item[SAS] Semi-analytical solution.
    \item[SOE] Second-order explicit scheme. 
    \item[TVD] Total variation decreasing.
\end{IEEEdescription}
\subsection{Operators}
\begin{IEEEdescription}[\IEEEusemathlabelsep\IEEEsetlabelwidth{$\mathrm{diag}(\cdot)$}]
    \item[$\mathrm{diag}(\cdot)$] Transform a column vector into a diagonal matrix.
    \item[$\otimes$] Convolution.
    \item[$\bX(0:k)$] Linear combination of $\bX(0), \bX(1),\cdots, \bX(k)$.
\end{IEEEdescription}
\subsection{Variables}
The capitalized/bold forms of these variables denote the DT/vector of these variables respectively. 
\begin{IEEEdescription}[\IEEEusemathlabelsep\IEEEsetlabelwidth{$\btau^\text{in/out}$}]
    \item[$\m/\mq$] Pipe/node injection mass flow rate.
    \item[$\tau^\text{s/r}$] Node supply/return temperature.
    \item[$\tau^\text{in/out,s/r}$] Inlet/outlet temperature of supply/return pipes.
    \item[$\tau^\mathrm{amb}$] Ambient Temperature.
    \item[$\phi$] Node power.
    \item[$e/f$] Real/imaginary part of bus voltage.
    \item[$p/q$] Active/reactive power.
\end{IEEEdescription}
\subsection{Coefficients}
\begin{IEEEdescription}[\IEEEusemathlabelsep\IEEEsetlabelwidth{$\Delta x/\Delta t$}]
\item[$\gamma$] Cross-sectional area.
\item[$\rho$] Density.
\item[$H$] Length of pipe.
\item[$C_\mathrm{p}$] Thermal capacity of mass flow.
\item[$\lambda$] Overall heat transfer coefficient.
\item[$K$] Resistance coefficient.
\item[$Z$] Trade-off between heat supply and electric power.
\item[$c_{\mathrm{m1}}$] Heat-to-power ratio.
\item[$\eta_\text{e}$] Electrical efficiency.
\item[$F_\text{in}$] Fuel input rate of steam turbine. 
\item[$\Delta x/\Delta t$] Spatial/temporal step sizes.
\item[$\theta$] Parameter that balances the dissipative/dispersive errors in PDE solutions.
    \item[$V/L$] Node/loop incidence matrix.
    \item[$G/B$] Conductance/susceptance matrix. 
\end{IEEEdescription}
\subsection{Superscripts/Subscripts}
\begin{IEEEdescription}[\IEEEusemathlabelsep\IEEEsetlabelwidth{PQ/PV/R}]
    \item[D] Diagonal matrix generated by $\mathrm{diag}(\cdot)$.
    \item[PQ/PV/R] Row indices of vector/matrix related to PQ/PV/Slack buses in EPSs.
    \item[R/S/L/I] Row indices of vector/matrix related to slack/source/load/intermediate nodes in DHSs.
    \item[SP] Specified value.
\end{IEEEdescription}
\section{Introduction}
\IEEEPARstart{I}{n} the recent decade, the applications of heat and electricity integrated energy systems(HE-IESs) have been prevailing. For one thing, the integration of district heating systems (DHSs) and electric power systems (EPSs) provides more flexibility and helps improve the economy\cite{lizhigang2016,lushuai2020,Zhang2021}; for another, state changes and outages are transferred from one system to the other, threatening the overall operation security\cite{Pan2016,Pan2019,zhangsuhan2021dynamic_security_control}. In view of this, quasi-dynamic energy flow calculation, which simulates the real operation scenarios and foresees the upcoming contingencies, becomes indispensable.
\par 
In HE-IESs, the aim of quasi-dynamic energy flow calculation is to obtain: 1) in DHS side, the variations of temperature distribution, heat power and mass flow, and 2) in EPS side, the variations of voltage and electric power. This is realized by solving the partial differential algebraic equations which are comprised of the EPS models and the DHS models\cite{Qin2019,zhangsuhan2021superposition}. Nonlinear power flow equations constitute the EPS models, whereas the complexity of DHS models depends on the regulation modes. In quality regulation mode, the operators fix the mass flow rates and regulate the DHS by adjusting source supply temperatures, which means the hydraulics are predetermined and remain unchanged. Hence, the thermal and energy algebraic equations, and the partial differential equations (PDEs) governing thermal dynamics in pipes are linear. These models are relatively easy to solve and their analytical solutions have been derived in \cite{Yangjingwei2020,chenyuwei2021,zhangsuhan2021fullyanalytical,Zhengjinfu2017}. 
In quantity regulation mode, operators fix the source supply temperatures and regulate the DHS by adjust mass flow rates. As a result, couplings between temperature and mass flow variables bring nonlinearity to DHS models. 
\par
Nonlinearity is intractable. Traditional iteration solvers cannot guarantee to find exact solutions of these nonlinear equations within prescribed number of steps. Also, in each of the iteration steps, highly dimensional Jacobian matrices are updated and factorized, which is computationally intensive. On the other hand, the convergence of iteration solvers on this complex systems have not been theoretically secured, so that divergence is a common issue. In view of this, this paper focuses on the nonlinear DHS models and aims to develop a new efficient and robust solver for the quasi-dynamic energy flow calculation in HE-IESs.
\par 
Typically, PDE in the HE-IES models are first converted into algebraic equations by finite difference methods (FDMs) with different accuracy and stability properties, which include: 1) implicit upwind (IU) scheme\cite{Wangyaran2017}; 2) the modified characteristic line methods\cite{Wangyaran2017,Denarie2019}; 3) Second-order Explicit(SOE) scheme\cite{Yao2021}. However, little attention is paid to the decayings of high-frequency components, which are called dissipative errors\cite{Thomas1995}, and the fake oscillations, which are called dispersive errors\cite{Thomas1995}, in solutions by these methods. It is later shown in case studies that the widely used IU scheme has distinct dissipative errors and the SOE scheme has distinct dispersive errors, which undoubtedly weaken the credibility of the quasi-dynamic energy flow calculation. 
Choosing proper temporal and spatial step sizes may help reduce these errors, but, as shown in Appendix \ref{paper DT appendix CFL Number, Dissipative and Dispersive Errors}, this becomes in vain when mass flow velocities are variable. 
To reduce the dissipative and dispersive errors regardless of the mass flow velocities and spatial/temporal step sizes, schemes with total variation decreasing (TVD) property have been developed in \cite{Thomas1995,AlexanderKurganov2000} and work well.
\par 
After applying FDMs to the PDEs, we should solve the consequent nonlinear algebraic equations which are composed of the discretized PDEs, the nonlinear thermal/hydraulic/energy equations and the nonlinear power flow equations. Reference \cite{Qin2019} has proposed the HE-FBI method to alternatively solve these equations until the convergence criteria are satisfied. However, the alternating solution strategy passes errors between different iteration loops. For example, the values of EPS variables used in DHS model calculation may have errors bigger than convergence criteria and would not be updated in the current iteration loop. These errors can be called alternating errors and increase the risk of divergence. In view of this, reference \cite{zhangsuhan2021partitional} and \cite{zhangsuhan2021superposition} have tried to improve the computation efficiency and convergence of the DHS side by uncoupling the DHSs into small simple sub-systems. Reference \cite{Massrur2018} has applied holomorphic embedding method to eliminate the iterations of the EPS side. However, iterations, which account for most of the computation overhead in quasi-dynamic energy flow calculation, have not been totally eliminated yet. Also, these methods are still prone to divergence because of alternating errors and improper initial guesses.
\par
Recently, differential transformation (DT) has been proposed to solve nonlinear power system models. This method derives time-polynomial solutions, which are called semi-analytical solutions (SASs), of electro-mechanical models\cite{Liu2019,Liu2020} and the modified continuation power flow models\cite{Liu2020_2}. This method demonstrates fast and reliable performance compared with traditional numerical-integration methods for the following two reasons: 1) it gets rid of iterations completely; 2) it constructs high-order approximations conveniently. These researches shed light on a new non-iterative solver of the nonlinear partial differential algebraic models of HE-IESs. However, DTs developed in \cite{Liu2019,Liu2020,Liu2020_2} can only cope with ordinary differential algebraic equations. Therefore, there is a need for converting the PDEs into ordinary differential equations (ODEs). 
\par
In this paper, we present a non-iterative algorithm for quasi-dynamic energy flow based on DT. The proposed method is able to derive SASs of all the unknown variables in the nonlinear models, but requires only fixed computation resources in each time window. Furthermore, the proposed method solves EPS and DHS models together and hence alternating errors are avoided. The kernel of the proposed method is to discretize the spatial derivative of the PDE only, which converts the PDE into ODEs. Then we can apply DT to the resulting ordinary differential algebraic equations of the spatially discretized HE-IES models. In this paper, we use a scheme with TVD property to realize the conversion, which reduces dissipative and dispersive errors effectively. Compared with \cite{wanghai2018} which has used the same TVD scheme, we achieve higher temporal accuracy because we use DT as the ODE solver instead of simple forward difference of temporal derivative. Additionally, to ensure calculation robustness and efficiency, we develop an easy-implemented adaptive time window control strategy based on the recursive nature of DT. 
    \par
    The major contributions are summarized as follows.
    \par
    1) A DT-based non-iterative algorithm is proposed to improve the efficiency and robustness of quasi-dynamic energy flow calculation in HE-IESs. With the aid of a semi-discrete TVD scheme, the dissipative and dispersive errors in thermal dynamics are effectively reduced.
    \par
    2) An adaptive time window control strategy is designed to further accelerate the simulation routines and to avoid non-convergence by selecting appropriate temporal step sizes.
\par 
This paper is outlined in the following way: Section \ref{paper DT section Introduction to the Proposed DT-based Method} introduces DT; Section \ref{paper DT section DTs of PDEs} presents the DT-based PDE solver; Section \ref{paper DT section DTs of HE-IESs} derives the DTs of nonlinear algebraic equations; Section \ref{paper DT section calculation procedure} illustrates the generalized DT-based SAS-derivation framework on a small system, introduces the adaptive time window control strategy and gives the overall pseudocode of the proposed method; Section \ref{paper DT section case studies} gives the case studies; Section \ref{paper DT section conclusion} concludes. 
\section{Introduce Differential Transformation}\label{paper DT section Introduction to the Proposed DT-based Method}
DT derives SASs of state variables of ODEs by obtaining the coefficients of their Taylor series.
Following \cite{Liu2019,Liu2020,Liu2020_2}, these coefficients are called DT coefficients and are defined as \eqref{paper DT equation definition DT_coeff}. The corresponding SAS can be written as \eqref{paper DT equation definition DT_AAS}. The $K+1$-term SAS \eqref{paper DT equation definition DT_AAS} is called DT-$K$ for short in this paper.
\begin{equation}
    \label{paper DT equation definition DT_coeff}
    \bX(k)=\frac{1}{k!}\left[\frac{\mathrm{d}^k\bx(t)}{\mathrm{d}t^k}\right]_{t=0}
\end{equation}
\begin{equation}\label{paper DT equation definition DT_AAS}
    \bx(t)=\sum_{k=0}^{K}\bX(k)\cdot t^k
\end{equation}
\par
To obtain $\bX(0)$ to $\bX(K)$, DT transforms the original nonlinear ODEs about $\bx(t)$ into linear recursive equations about $\bX(k)$ with the following rules, where we denote by $\bY(k)$ the DT coefficient of $\by(t)$; $c\in \mathbb{R}$; $\bx(t), \by(t), \bX(k), \bY(k), \boldsymbol{\delta}(k) \in \mathbb{R}^{n\times 1}$; $\bX_\mathrm{D}(k)\in \mathbb{R}^{n\times n}$; $\md(\cdot)$ is an operator which transforms a column vector into a diagonal matrix.
\par
\begin{enumerate}
    \item $\bx(0)\rightarrow \bX(0)$.
    \item $c\bx(t)\rightarrow c\bX(k)$.
    \item $\bx(t)\pm \by(t)\rightarrow \bX(k)\pm \bY(k)$.
    \item $\md(\bx(t))\by(t)\rightarrow $
    \[\bX_\mathrm{D}(k)\otimes \bY(k)=\sumnol_{m=0}^k\bX_\mathrm{D}(m)\bY(k-m).\]
    \item $c\rightarrow c\boldsymbol{\delta} (k)$, where
    \begin{equation*}
        \boldsymbol{\delta} (k)=\left\{
        \begin{aligned}
             & \boldsymbol{1},\quad k=0     \\
             & \boldsymbol{0},\quad k\neq 0
        \end{aligned}\right..
    \end{equation*}
    \item $\mathrm{d}\bx(t)/\mathrm{d}t\rightarrow (k+1)\bX(k+1)$.
\end{enumerate}
\par
The readers can refer to \cite[\S I.8]{wanner1993solving,Liu2019,Liu2020,Liu2020_2} for proofs and rules for more complex functions. The next section can be viewed as a first tutorial of DT.
\section{Discretize and Derive DT of PDE}\label{paper DT section DTs of PDEs}
The PDE governing thermal dynamics is essentially a one-dimensional hyperbolic convection equation with source term\cite{Qin2019}, which is
\begin{equation}
    \frac{\partial\tau}{\partial t}+\frac{\dot{m}}{\gamma \rho}\frac{\partial\tau}{\partial x}+\frac{\lambda }{\gamma\rho C_\mathrm{p}}(\tau-\tau^\mathrm{amb})=0,\label{paper DT equation PDE heat pipe model}
\end{equation}
    where $\tau$ is the two-dimensional distribution of temperature along time $t$ and position $x$; $\m$ is the mass flow rate; $\gamma$ is the cross-sectional area of the pipe; $\rho$ is the density of mass flow; $\lambda$ is the overall hear transfer coefficient; $C_\mathrm{p}$ is the thermal capacity of mass flow; $\tau^\text{amb}$ is the ambient temperature,
with initial condition
\begin{equation}
    \tau(x,0)=\varphi (x),\quad x\geq 0,\label{paper DT initial condition}
\end{equation}
and boundary condition
\begin{equation}
    \tau(0,t)=\psi (t),\quad t\geq 0.\label{paper DT boundary condition}
\end{equation}
\par
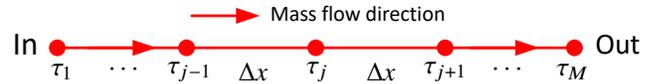
\begin{figure}[!h]
    \centering
    \begin{tikzpicture}
        \node[anchor=south west,inner sep=0] at (0,0) {\includegraphics[width=3.3in]{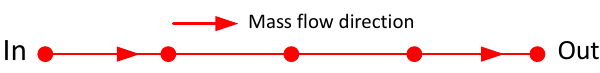}};
        \node (T1) at (0.7,-0.05) {$\tau_1$};
        \node (dot1) at (1.55,-0.05) {$\cdots$};
        \node (T2) at (2.4,-0.05) {$\tau_{j-1}$};
        \node (h1) at (3.25,-0.05) {$\Delta x$};
        \node (T3) at (4.1,-0.05) {$\tau_j$};
        \node (h2) at (4.95,-0.05) {$\Delta x$};
        \node (T4) at (5.8,-0.05) {$\tau_{j+1}$};
        \node (dot2) at (6.65,-0.05) {$\cdots$};
        \node (T5) at (7.5,-0.05) {$\tau_M$};
    \end{tikzpicture}\\
    \caption{Semi-discrete difference of pipe.}\label{paper DT figure semi-discrete}
\end{figure}

 
Semi-discrete difference scheme is widely applied to solving the hyperbolic and parabolic PDEs\cite{AlexanderKurganov2000,MedovikovAA1998}. By replacing the spatial derivative $\partial\tau/\partial x$ by its discrete approximation, the PDEs are converted into ODEs, which can be solved by ODE solvers such as Runge-Kutta and Euler methods. In the simplest case, we can replace $\partial\tau/\partial x$ by the backward difference quotient $(\tau_{j}-\tau_{j-1})/\Delta x$ where $\Delta x$ is the spatial step size. If we assume that total length of the pipe is $H$, then the PDE becomes $M-1$ ODEs about temperatures at each discrete node as shown in Fig. \ref{paper DT figure semi-discrete} where $M=H/\Delta x+1$. $\tau_1$ corresponds to the boundary conditions and is viewed as a known variable.
\par 
However, semi-discrete difference scheme based on backward difference quotient suffers dissipative and dispersive errors.
In this paper, we discretize the spatial derivative of \eqref{paper DT equation PDE heat pipe model} with the semi-discrete TVD scheme proposed in \cite{AlexanderKurganov2000}, which derives
\begin{equation}
    \begin{aligned}
    \frac{\mathrm{d}\tau_j}{\mathrm{d}t}=&\frac{\dot{m}}{\gamma \rho\Delta x}\left(\tau_{j-1}+\frac{\Delta x}{2}(\tau_{j-1})_x-\tau_{j}-\frac{\Delta x}{2}(\tau_j)_x\right) \\ -&\frac{\lambda }{\gamma\rho C_\mathrm{p}}(\tau_j-\tau^\mathrm{amb}),\quad 2\leq j\leq M
    \end{aligned}\label{paper DT equation time domain TVD scheme}
\end{equation}
where $(\tau_j)_x$ denotes the spatial derivative of $\tau_{j}$ and its formula is not distinct but dynamically decided by minmod slope limiter
\begin{equation}\label{paper DT equation minmod slope limiter}
    \mathrm{minmod}(\chi_1,\chi_2,\chi_3)=
    \begin{cases}
        \displaystyle\min(\chi_1,\chi_2,\chi_3) & \text{if }\forall \chi_i>0,\\
        \displaystyle\max(\chi_1,\chi_2,\chi_3) & \text{if }\forall \chi_i<0,\\
        0 & \mathrm{otherwise} . \\
    \end{cases}
\end{equation}
where 
\[
    \begin{aligned}
        &\chi_1=\theta\frac{\tau_j(0)-\tau_{j-1}(0)}{\Delta x},\\ 
        &\chi_2=\frac{\tau_{j+1}(0)-\tau_{j-1}(0)}{2\Delta x},\\ 
        &\chi_3=\theta\frac{\tau_{j+1}(0)-\tau_{j}(0)}{\Delta x}.        
    \end{aligned}
\]
$\theta\in [1,2]$ balances the dissipative and dispersive errors in the solutions. Choices of $\theta$ are later studied in Section \ref{paper DT section case studies}. $\tau_j(0)$ denotes the initial value of $\tau_j$ in the current time window. We put
\[
    (\tau_j)_x=    
    \begin{cases}
        \theta\frac{\tau_j-\tau_{j-1}}{\Delta x}& \text{if }\mathrm{minmod}(\chi_1,\chi_2,\chi_3)=\chi_1,\\
        \frac{\tau_{j+1}-\tau_{j-1}}{2\Delta x}& \text{if }\mathrm{minmod}(\chi_1,\chi_2,\chi_3)=\chi_2,\\
        \theta\frac{\tau_{j+1}-\tau_{j}}{\Delta x}&\text{if }\mathrm{minmod}(\chi_1,\chi_2,\chi_3)=\chi_3,\\
        0 & \mathrm{otherwise} . \\
    \end{cases}
\]
In each time window, we should first give a distinct formula to each $(\tau_j)_x$ according to the output of minmod slope limiter. Then \eqref{paper DT equation time domain TVD scheme} becomes ODE about $\tau_{j-2}$, $\tau_{j-1}$, $\tau_{j}$ and $\tau_{j+1}$. For the $j=2$ and $j=M$ cases where $\tau_0$ and $\tau_{M+1}$ do not exist, we  assumes $(\tau_j)_x$ and $(\tau_{j-1})_x$ to be zero in \eqref{paper DT equation time domain TVD scheme}.
\par
The scheme \eqref{paper DT equation time domain TVD scheme} will be of second-order spatial accuracy only in the cases where $(\tau_j)_x=(\tau_{j+1}-\tau_{j-1})/({2\Delta x})$ and $(\tau_{j-1})_x=(\tau_{j}-\tau_{j-2})/({2\Delta x})$ and these cases usually occur on smooth sections of the solutions. Otherwise, it will be spatially first-order accurate and these cases usually occur in discontinuous sections of solutions.
\par 
Here, we use DT to solve these ODEs and exemplify the derivation of DT of \eqref{paper DT equation time domain TVD scheme} by assuming $j=2$ or $M$. The cases where $2<j<M$  can be derived in the same way.
\par 
In quantity regulation mode, $\dot{m}$ is also variable.
Applying transformation rule 6) to left hand side and 2)3)4)5) to right hand side of
\begin{equation*}
    \frac{\mathrm{d}\tau_j}{\mathrm{d}t}=\frac{\dot{m}}{\gamma \rho\Delta x}(\tau_{j-1}-\tau_{j})-\frac{\lambda }{\gamma\rho C_\mathrm{p}}(\tau_j-\tau^\mathrm{amb})     
\end{equation*}
derives
\begin{equation}
    \begin{aligned}
        (k+1)T_j(k+1)&=\frac{1}{\gamma \rho\Delta x}\dot{M}(k)\otimes(T_{j-1}(k)-T_{j}(k))\\
        &-\frac{\lambda }{\gamma\rho C_\mathrm{p}}(T_j(k)-\tau^\mathrm{amb}\delta(k))\\ 
    \end{aligned}\label{paper DT equation DT of PDE}
\end{equation}
where $T_j(k)$ is the DT of $\tau_j$; $\dot{M}(k)$ is the DT of $\m$. 
\par 
After the semi-discrete difference of PDE \eqref{paper DT equation PDE heat pipe model}, the initial condition becomes
\begin{equation}\label{paper DT equation initial condition after semi-discrete difference}
    \tau_j(0)=\varphi (x_j),\quad 2\leq j\leq M,
\end{equation}
where $x_j$ is the position of discrete node $j$, and the boundary condition becomes
\begin{equation}\label{paper DT equation boundary condition after semi-discrete difference}
    \tau_1(t)=\psi(t).
\end{equation}
\par
According to transformation rule 1), we obtain
\begin{equation*}
    T_j(0)=\varphi (x_j),\quad 1\leq j\leq M.
\end{equation*}
\par 
As for boundary condition $\psi (t)$ with distinct formula such as polynomials, trigonometric functions, etc., we should first obtain the recursive expression of the DT of $\psi (t)$, which is denoted by $\Psi(k)$. Then derive DT of \eqref{paper DT equation boundary condition after semi-discrete difference}, we obtain
\begin{equation*}
     T_1(k)=\Psi (k),\quad k\geq 0.
\end{equation*}
\par  It is obvious that right hand side of \eqref{paper DT equation DT of PDE} only contains DT coefficients up to order $k$. If we are given the variation of $\m$, we can derive $\M(0)$ to $\M(K)$ first and then obtain, in the following order,
\begin{enumerate}
    \item $T_j(1)$ by $\M(0)$ and $T_i(0)$;
    \item $T_j(2)$ by $\M(0)$ to $\M(1)$ and $T_i(0)$ to $T_i(1)$;
    \item $T_j(3)$ by $\M(0)$ to $\M(2)$ and $T_i(0)$ to $T_i(2)$;
    \item ...
    \item $T_j(K)$ by $\M(0)$ to $\M(K-1)$ and $T_i(0)$ to $T_i(K-1)$;
\end{enumerate}where $2\leq j\leq M$, $1\leq i\leq M$ and $T_1(0)$ to $T_1(K)$ are given by the boundary condition.
\par 
After obtaining $T_j(0)$ to $T_j(K)$, we can obtain the time polynomial depicting temperature variation as 
\begin{equation*}
    \tau_j(t)=\sum_{k=0}^{K}T_j(k)\cdot t^k.
\end{equation*}
\par 
In this section, we apply DT to the PDE of a single pipe whose initial-boundary conditions and mass flow variation $\dot{m}$ are known. However, in HE-IESs, boundary conditions of pipes and $\m$ are not explicitly given. In following sections, we will derive the DT of HE-IES models first and illustrate how to deal with this case in Section \ref{paper DT section calculation procedure1}.
\section{Derive DTs of Nonlinear Algebraic Equations}\label{paper DT section DTs of HE-IESs}
\subsection{DTs of DHS Models}
This paper considers the quantity regulation mode of DHSs and hence follows the DHS models and the node type assumptions adopted by \cite{Qin2019,zhangsuhan2021partitional,Liuxuezhi2016}. The known and unknown variables of DHS nodes are shown in Table \ref{paper DT table knowns and unknowns} where $\phi$/$\taus$/$\taur$/$\mq$ are respectively heat power/node supply temperature/node return temperature/node injection mass flow rate. 
\begin{table}[!h]
    \centering
    \caption{Node Type Assumption in DHSs}\label{paper DT table knowns and unknowns}
    \begin{tabular}{ccccc}
        \hline
        Node Type&Slack (R)&Source (S)&Load (L)&Intermediate (I)\\
        \hline
        Known&$\taus$&$\phi, \taus$&$\phi, \taur$&$\phi= \mq=0$\\
        Unknown&$\phi, \taur, \mq$&$\taur, \mq$&$\taus, \mq$&$\taus, \taur$\\
        \hline
    \end{tabular}   
\end{table}
\par
Below, we use subscript R, S, L and I to denote the row indices of matrices or vectors related to slack, source, load and intermediate nodes, respectively. The combination of these subscripts denotes the combination of these nodes. For example, subscript {R,S} denotes the combination of slack and source nodes.
\subsubsection{Mass Flow Continuity Equations}
node injection mass flow rate is equal to the mass flow rate that enters into the node minus the mass flow rate leaves from the node, which yields
\begin{equation}
    V_\text{R,S}\bm=-\bmq_\text{R,S},\label{paper DT AE Mass Flow Continuity Equations1}
\end{equation}
\begin{equation}
    V_\text{L}\bm=\bmq_\text{L},\label{paper DT AE Mass Flow Continuity Equations2}
\end{equation}
\begin{equation}
    V_\text{I}\bm=0,\label{paper DT AE Mass Flow Continuity Equations3}
\end{equation}
where $V$ is the node incidence matrix and its element
\[
V_{ij}=\left\{ \begin{aligned}
    1,&\quad \text{if pipe}\ j\ \text{flows into node}\ i\\
    -1,&\quad \text{if pipe}\ j\ \text{flows out of node}\ i\\ 
    0,&\quad \text{otherwise}
\end{aligned}\right..
\]
$V$ describes the supply networks. In this paper, we consider the case where the mass flow in supply and return networks are symmetric, so we only need to study the hydraulic equations of supply networks here. $\bm$ is the vector of mass flow rate; $\bmq$ is the vector of node injection mass flow rate.
There is a minus sign before $\bmq_\text{R,S}$ to ensure that node injection mass flow rates of slack and source nodes are negative. 
\par 
Applying transformation rule 2) to both sides of \eqref{paper DT AE Mass Flow Continuity Equations1}, \eqref{paper DT AE Mass Flow Continuity Equations2} and \eqref{paper DT AE Mass Flow Continuity Equations3} derives
\begin{equation}
   V_\text{R,S}\bM(k)=-\bMq_\text{R,S}(k), \label{paper DT k domain AE Mass Flow Continuity Equations1}
\end{equation}
\begin{equation}
    V_\text{L}\bM(k)=\bMq_\text{L}(k), \label{paper DT k domain AE Mass Flow Continuity Equations2}
\end{equation}
\begin{equation}
    V_\text{I}\bM(k)=0, \label{paper DT k domain AE Mass Flow Continuity Equations3}
\end{equation}
where $\bM(k)$ and $\bMq(k)$ are DTs of $\bm$ and $\bmq$. \eqref{paper DT k domain AE Mass Flow Continuity Equations1}, \eqref{paper DT k domain AE Mass Flow Continuity Equations2} and \eqref{paper DT k domain AE Mass Flow Continuity Equations3} are linear equations for $\bM(k)$ and $\bMq(k)$ because the coefficients of them are elements of node incidence matrix, which are constant. 
\subsubsection{Loop Pressure Equations}
The loop pressure equation states that the sum of head losses around a closed loop must be equal to zero, which is
\begin{equation}
    L\md(K)\md(\bm)\bm=\boldsymbol{0},\label{paper DT AE Loop Pressure Equations}
\end{equation}
where $L$ is the loop incidence matrix and its element
\[
    L_{ij}=
    \left\{ 
    \begin{aligned}
        1,&\quad \text{if pipe}\ j\ \text{in loop}\ i\ \text{is clockwise}\\
        -1,&\quad \text{if pipe}\ j\ \text{in loop}\ i\ \text{is counterclockwise}\\ 
        0,&\quad \text{if pipe}\ j\ \text{is not in loop}\ i
    \end{aligned}\right.;
\]
$K$ is the vector of pipe resistance coefficient. The same as $V$, $L$ also describes the supply networks.
\par Applying transformation rules 2) and 4) to \eqref{paper DT AE Loop Pressure Equations} derives
\begin{equation}
    LK_\text{D}\bM_\text{D}(k)\otimes \bM(k)=\boldsymbol{0}\label{paper DT k-domain AE Loop Pressure Equations}
\end{equation}
where the subscript D denotes the diagonal matrix transformed by $\md(\cdot)$. The coefficient of $\bM(k)$ is a constant coefficient multiple of $\bM(0)$ while $\bM(0)$ is the initial value of $\bm$.
Hence, the nonlinear equation \eqref{paper DT AE Loop Pressure Equations} for $\bm$ is transformed into the linear equation \eqref{paper DT k-domain AE Loop Pressure Equations} for $\bM(k)$.
\subsubsection{Node Temperature Equations}
Node temperature equals the mixture of pipe outlet temperatures, which, according to Table \ref{paper DT table knowns and unknowns}, yields
\begin{equation}\label{paper DT AE node temperature equations s mixture}
    \md(\btau_\text{L,I}^\text{s})V_\text{L,I}^\text{+}\bm=V_\text{L,I}^\text{+}\md(\btau^\text{out,s})\bm,
\end{equation}
\begin{equation}\label{paper DT AE node temperature equations r mixture}
    \md(\btau_\text{R,S,I}^\text{r})V_\text{R,S,I}^\text{-}\bm=V_\text{R,S,I}^\text{-}\md(\btau^\text{out,r})\bm,
\end{equation}
where $\btau^\text{s/r}$ is the vector of node supply/return temperatures, $\btau^\text{out,s/r}$ is the vector of the outlet temperatures of supply/return pipes; $V^{+}=\max(\boldsymbol{0},V)$ while $V^{-}=\min(\boldsymbol{0},V)$. Notation explanations to \eqref{paper DT AE node temperature equations s mixture} and \eqref{paper DT AE node temperature equations r mixture} can be found in Appendix \ref{paper DT appendix Notation Explanation}.
\par 
Applying transformation rules 2) and 4) to \eqref{paper DT AE node temperature equations s mixture} and \eqref{paper DT AE node temperature equations r mixture} respectively derives 
\begin{equation}
    \bT_\text{L,I,D}^{\text{s}}(k)\otimes  V_\text{L,I}^\text{+}\bM(k)= V_\text{L,I}^\text{+}\bT_\text{D}^{\text{out,s}}(k)\otimes \bM(k),\label{paper DT k-domain AE node temperature equations s}
\end{equation}
\begin{equation}
    \bT_\text{R,S,I,D}^{\text{r}}(k)\otimes  V_\text{R,S,I}^\text{-}\bM(k)= V_\text{R,S,I}^\text{-}\bT_\text{D}^{\text{out,r}}(k)\otimes \bM(k),\label{paper DT k-domain AE node temperature equations r}
\end{equation}
where $\bT^{\text{s/r}}(k)$ is the DT of $\btau^\text{s/r}$; $\bT^{\text{out,s/r}}(k)$ is the DT of $\btau^\text{out,s/r}$. In \eqref{paper DT k-domain AE node temperature equations s}, the coefficients of $\bM(k)$ are linear combinations of elements of $\bT^{\text{s}}(0)$ and $\bT^{\text{out,s}}(0)$ while the coefficients of elements of $\bT^{\text{s}}(k)$ and $\bT^{\text{out,s}}(k)$ are $\bM(0)$. Hence, the nonlinear equation \eqref{paper DT AE node temperature equations s mixture} about $\bm$, $\btau_\text{L,I}^\text{s}$ and $\btau^\text{out,s}$ is transformed into the linear equation \eqref{paper DT k-domain AE node temperature equations s} about $\bM(k)$, $\bT_\text{L,I}^{\text{s}}(k)$ and $\bT^{\text{out,s}}(k)$. Similar observation also applies to \eqref{paper DT k-domain AE node temperature equations r}.
\subsubsection{Node Power Equations}
Heat power produced or consumed at slack/source/load nodes satisfies
    \begin{equation}\label{paper DT AE node power equations}
        \bphi_\text{R,S,L}=C_\text{p}\md(\bmq_\text{R,S,L})(\btau_\text{R,S,L}^\text{s}-\btau_\text{R,S,L}^\text{r})
    \end{equation}
    where $\bphi$ is the vector of node power. 
    \par 
    Applying transformation rules 2) and 4) to right hand side of \eqref{paper DT AE node power equations} derives
    \begin{equation}\label{paper DT k-domain AE node power equations}
    \bPhi_\text{R,S,L}(k)=C_\text{p}\bMq_\text{R,S,L,D}(k)\otimes(\bTs_\text{R,S,L}(k)-\bTr_\text{R,S,L}(k)) 
    \end{equation}
    Likewise, the nonlinear equation \eqref{paper DT AE node power equations} about $\bphi$, $\bmq$, $\btau^\text{s}$ and $\btau^\text{r}$ are transformed into the linear equation \eqref{paper DT k-domain AE node power equations} about $\bPhiast(k)$, $\bMqast(k)$, $\bTsast(k)$ and $\bTrast(k)$, the coefficients of which are linear combinations of constants, $\bTs(0)$, $\bTr(0)$ and $\bMq(0)$.
\subsection{DTs of EPS Models}
The power flow models are
\begin{equation}\label{paper DT AE EPS power flow equations1}
    \bp=\md(\bee) (G\bee-B\bff)+\md (\bff)(B\bee+G\bff),        
\end{equation}
\begin{equation}\label{paper DT AE EPS power flow equations2}
    \bq=\md(\bff) (G\bee-B\bff)-\md (\bee)(B\bee+G\bff),        
\end{equation}
\begin{equation}\label{paper DT AE EPS power flow equations3}
    \md(\bee_\text{PV})\bee_\text{PV}+\md(\bff_\text{PV})\bff_\text{PV}=\md(\bU_\text{PV}^\mathrm{SP})\bU_\text{PV}^\mathrm{SP},
\end{equation}
\begin{equation}\label{paper DT AE EPS power flow equations4}
    e_\text{R}=e^\text{SP},\quad f_\text{R}=f^\text{SP},
\end{equation}
where $\bp/\bq$ are respectively the vectors of active/reactive power; $\bee/\bff$ are respectively the vectors of real/imaginary part of bus voltage; $G/B$ are respectively the conductance/susceptance matrices; $\bU$ is the vector of bus voltage magnitude; subscript PV/R denotes the indices of rows in matrices/vectors related to PV/slack buses; superscript SP denotes that the values are specified. We denote by $\bP(k)/\bQ(k)/\bE(k)/\bF(k)$ the DT of $\bp/\bq/\bee/\bff$ respectively.
\par
DTs of \eqref{paper DT AE EPS power flow equations1} to \eqref{paper DT AE EPS power flow equations4} have been derived in \cite{Liu2020_2} using a set of different notations, which are
\begin{equation}
    \begin{aligned}
    \bP(k)=&\bE_\text{D}(k)\otimes(G\bE(k)-B\bF(k))\\
    +&\bF_\text{D}(k)\otimes(B\bE(k)+G\bF(k)).\label{paper DT k domain AE EPS power flow equations1}
    \end{aligned}
\end{equation}
\eqref{paper DT k domain AE EPS power flow equations1} is a linear equation about $\bP(k)$, $\bE(k)$, $\bF(k)$, the coefficients of which are linear combinations of constants, $\bE(0)$ and $\bF(0)$.
\par 
Likewise, we have the DTs of \eqref{paper DT AE EPS power flow equations2}, \eqref{paper DT AE EPS power flow equations3} and \eqref{paper DT AE EPS power flow equations4}, which are linear equations about $\bQ(k)$, $\bE(k)$ and $\bF(k)$ as follows.
\begin{equation}
    \begin{aligned}
    \bQ(k)=&\bF_\text{D}(k)\otimes(G\bE(k)-B\bF(k))\\
    -&\bE_\text{D}(k)\otimes(B\bE(k)+G\bF(k)).\label{paper DT k domain AE EPS power flow equations2}
    \end{aligned}
\end{equation}
\begin{equation}
    \bE_\text{PV,D}(k)\otimes\bE_\text{PV}(k)+\bF_\text{PV,D}(k)\otimes\bF_\text{PV}(k)=\bU_\text{PV,D}^\mathrm{SP}\bU_\text{PV}^\mathrm{SP}\delta(k).\label{paper DT k domain AE EPS power flow equations3}
\end{equation}
\begin{equation}
    \bE_\text{R}(k)=e^\text{SP}\delta(k),\quad \bF_\text{R}(k)=f^\text{SP}\delta(k).\label{paper DT k domain AE EPS power flow equations4}
\end{equation}
\subsection{DTs of Coupling Components}
According to \cite{Liuxuezhi2016}, the relationships between heat and electric power of extraction steam turbine and gas turbine respectively satisfy
\begin{equation}
    \bp=-\bphi/Z+\eta_\mathrm{e}F_\mathrm{in},\label{paper DT AE Coupling Components1}
\end{equation}
\begin{equation}
    \bphi=c_{\mathrm{m1}}\bp,\label{paper DT AE Coupling Components2}
\end{equation}
where $Z$ is the ratio that describes the trade-off between heat supplied to the site and the electric power of the extraction steam turbine; $c_{\mathrm{m1}}$ is the heat-to-power ratio of the gas turbine; $\eta_\text{e}$ is the electrical efficiency; $F_\text{in}$ is the fuel input rate of the steam turbine. 
\par 
Applying transformation rules 2), 3) and 5) to \eqref{paper DT AE Coupling Components1} and \eqref{paper DT AE Coupling Components2} derives 
\begin{equation}
    \bP(k)=-\bPhi(k)/Z+\eta_\mathrm{e}F_\mathrm{in}\boldsymbol{\delta}(k),\label{paper DT k domain AE Coupling Components1}
\end{equation}
\begin{equation}
    \bPhi(k)=c_{\mathrm{m1}}\bP(k).\label{paper DT k domain AE Coupling Components2}
\end{equation}
which are linear equations about $\bP(k)$ and $\bPhi(k)$.
\section{Illustrate SAS-derivation Framework and Introduce Adaptive Time Window Control}
\label{paper DT section calculation procedure}
\subsection{Illustrating the SAS-derivation framework on a small HE-IES}\label{paper DT section calculation procedure1}
\begin{figure}[!h]
    \centering
    \includegraphics[width=3.4in]{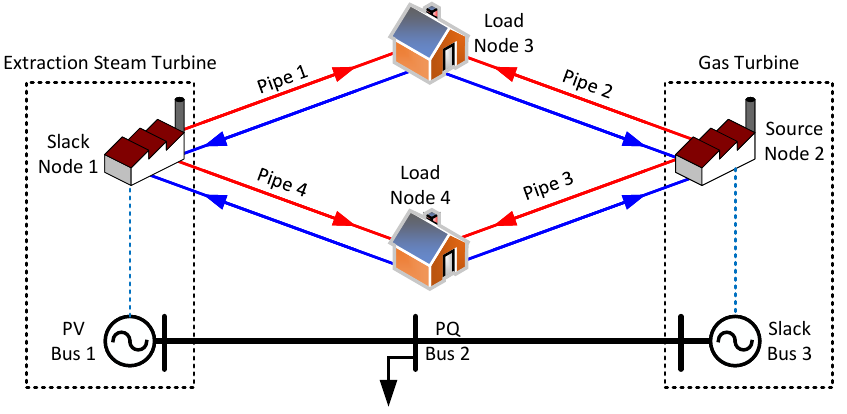}
    \caption{A small HE-IES with four DHS nodes and three EPS buses.}\label{paper DT figure small system}
\end{figure}
The proposed method is illustrated on the small system shown in Fig. \ref{paper DT figure small system} which is comprised of four DHS nodes and three EPS buses. The node and loop incidence matrices of the DHS are respectively
\[V=
\begin{bmatrix}
-1&0&0&-1\\ 
0&-1&-1&0\\ 
1&1&0&0\\ 
0&0&1&1
\end{bmatrix}\]
and
\[L=
\begin{bmatrix}
1&-1&1&-1\\ 
\end{bmatrix}.\]
\par 
We assume the spatial step size $\Delta x$ to be the length of each pipe. Therefore, PDE of each supply pipe is converted into an ODE about $\tauouts$ and $\tauins$ while PDE of each return pipe is converted into an ODE about $\tauoutr$ and $\tauinr$. We denote by $\btauinsr$ the vector of the inlet temperatures of supply/return pipes. The known and unknown variables in the HE-IES are shown in Table \ref{paper DT table knowns and unknowns in the illustrative EPS} and \ref{paper DT table knowns and unknowns of DHS nodes}. There are totally thirty-eight unknown variables. Different from Table \ref{paper DT table knowns and unknowns} and the bus type assumption in power flow model, $p_1$ is unknown because the electrical output of bus 1 is constrained by the heat output, which is $\phi_1$, of the extraction steam turbine and $\phi_1$ is unknown. $\phi_2$ is unknown because the heat output of node 2 is constrained by the electrical output, which is $p_3$, of the gas turbine and $p_3$ is unknown.
\begin{table}
    \centering
    \caption{Known and Unknown Variables of EPS}\label{paper DT table knowns and unknowns in the illustrative EPS}
    \begin{tabular}{cccc}
        \hline
        Bus Number&1&2&3\\
        \hline
        Type&PV&PQ&Slack\\
        Known&-&$p_2,q_2$&$e_3,f_3$\\
        Unknown&$e_1,f_1,p_1,q_1$&$e_2,f_2$&$p_3,q_3$\\
        \hline
    \end{tabular}
\end{table}
\begin{table*}
    \centering
    \caption{Known and Unknown Variables of DHS}\label{paper DT table knowns and unknowns of DHS nodes}
    \begin{tabular}{ccccccccc}
        \hline
        &Node 1&Node 2&Node 3&Node 4&Pipe 1&Pipe 2&Pipe 3&Pipe 4\\
        \hline
        Type&Slack&Source&Load&Load&-&-&-&-\\
        Known&$\taus_1$&$ \taus_2$&$\phi_3, \taur_3$&$\phi_4,\taur_4$&-&-&-&-\\
        Unknown&$\phi_1, \taur_1, \mq_1$&$\phi_2,\taur_2, \mq_2$&$\taus_3, \mq_3$&$\taus_4, \mq_4$&$\tauoutsr_1,\tauinsr_1,\m_1$&$\tauoutsr_2,\tauinsr_2,\m_2$&$\tauoutsr_3,\tauinsr_3,\m_3$&$\tauoutsr_4,\tauinsr_4,\m_4$\\
        \hline
    \end{tabular}   
\end{table*}
\begin{figure*}
    \begin{equation}\label{paper DT equation k-domain illustrative active power equation}
        \begin{aligned}
            \begin{bmatrix}
                P_1(1)\\ 
                P_2(1)\\ 
                P_3(1)
            \end{bmatrix}
            =&\md
            \begin{bmatrix}
                E_1(1)\\ 
                E_2(1)\\ 
                E_3(1)
            \end{bmatrix}
            \left(
            G
            \begin{bmatrix}
                E_1(0)\\ 
                E_2(0)\\
                E_3(0)
            \end{bmatrix}
            -
            B 
            \begin{bmatrix}
                F_1(0)\\ 
                F_2(0)\\
                F_3(0)
            \end{bmatrix}   
            \right)
            +\md
            \begin{bmatrix}
                F_1(1)\\ 
                F_2(1)\\ 
                F_3(1)
            \end{bmatrix}
            \left(
            B
            \begin{bmatrix}
                E_1(0)\\ 
                E_2(0)\\
                E_3(0)
            \end{bmatrix}
            +
            G 
            \begin{bmatrix}
                F_1(0)\\ 
                F_2(0)\\
                F_3(0)
            \end{bmatrix}   
            \right)
            \\
            +&\md 
            \begin{bmatrix}
                E_1(0)\\ 
                E_2(0)\\ 
                E_3(0)
            \end{bmatrix}
            \left(
            G
            \begin{bmatrix}
                E_1(1)\\ 
                E_2(1)\\
                E_3(1)
            \end{bmatrix}
            -
            B  
            \begin{bmatrix}
                F_1(1)\\ 
                F_2(1)\\
                F_3(1)
            \end{bmatrix}   
            \right)   
            +\md 
            \begin{bmatrix}
                F_1(0)\\ 
                F_2(0)\\ 
                F_3(0)
            \end{bmatrix}
            \left(
            B
            \begin{bmatrix}
                E_1(1)\\ 
                E_2(1)\\
                E_3(1)
            \end{bmatrix}
            +
            G 
            \begin{bmatrix}
                F_1(1)\\ 
                F_2(1)\\
                F_3(1)
            \end{bmatrix}   
            \right),
        \end{aligned}
    \end{equation}
    \begin{equation}\label{paper DT equation k-domain illustrative reactive power equation}
        \begin{aligned}
            \begin{bmatrix}
                Q_1(1)\\ 
                Q_2(1)\\ 
                Q_3(1)
            \end{bmatrix}
            =&\md
            \begin{bmatrix}
                F_1(1)\\ 
                F_2(1)\\ 
                F_3(1)
            \end{bmatrix}
            \left(
            G
            \begin{bmatrix}
                E_1(0)\\ 
                E_2(0)\\
                E_3(0)
            \end{bmatrix}
            -
            B  
            \begin{bmatrix}
                F_1(0)\\ 
                F_2(0)\\
                F_3(0)
            \end{bmatrix}   
            \right)
            -\md
            \begin{bmatrix}
                E_1(1)\\ 
                E_2(1)\\ 
                E_3(1)
            \end{bmatrix}
            \left(
            B
            \begin{bmatrix}
                E_1(0)\\ 
                E_2(0)\\
                E_3(0)
            \end{bmatrix}
            +
            G  
            \begin{bmatrix}
                F_1(0)\\ 
                F_2(0)\\
                F_3(0)
            \end{bmatrix}   
            \right)
            \\
            +&\md
            \begin{bmatrix}
                F_1(0)\\ 
                F_2(0)\\ 
                F_3(0)
            \end{bmatrix}
            \left(
            G
            \begin{bmatrix}
                E_1(1)\\ 
                E_2(1)\\
                E_3(1)
            \end{bmatrix}
            -
            B  
            \begin{bmatrix}
                F_1(1)\\ 
                F_2(1)\\
                F_3(1)
            \end{bmatrix}   
            \right)  
            - 
            \md
            \begin{bmatrix}
                E_1(0)\\ 
                E_2(0)\\ 
                E_3(0)
            \end{bmatrix}
            \left(
            B
            \begin{bmatrix}
                E_1(1)\\ 
                E_2(1)\\
                E_3(1)
            \end{bmatrix}
            +
            G  
            \begin{bmatrix}
                F_1(1)\\ 
                F_2(1)\\
                F_3(1)
            \end{bmatrix}   
            \right),
        \end{aligned}
    \end{equation}
    \begin{equation}\label{paper DT equation k-domain illustrative voltage equation}
        2E_1(0)E_1(1)+2F_1(0)F_1(1)=0.
    \end{equation}

    \end{figure*}
\par
First, we obtain the zeroth-order DT coefficients, which are the initial values of variables, by steady-state energy flow calculation\cite{Liuxuezhi2016}. Then we should derive the DT of known variables, for example, $\Ts_1(0)$ to $\Ts_1(K)$ of $\taus_1$. Next, we can start calculating the first-order DT coefficients of the above thirty-eight unknown variables based on the following three steps.
\par 
\emph{Step 1:} Calculate elements of vector $\bTouts(1)$ and $\bToutr(1)$ by \eqref{paper DT equation DT of PDE}. For example,
\[
        \begin{aligned}
                \Touts_1(1)&=\frac{1}{\gamma \rho \Delta x}\M(0)(\Tins_1(0)-\Touts_1(0)) \\
                &-\frac{\lambda}{\gamma\rho C_\mathrm{p}}(\Touts_1(0)-\tau^\mathrm{amb}\delta(0)/\tau_\mathrm{b})\\
        \end{aligned}
\]
where the right hand side are all known. The other seven unknown first-order DT coefficients can be calculated directly likewise. That is, there remain thirty unknown first-order DT coefficients.
\par 
\emph{Step 2:} By \eqref{paper DT k domain AE Mass Flow Continuity Equations1} and \eqref{paper DT k domain AE Mass Flow Continuity Equations2}, the mass flow continuity equations are transformed into four linear equations about $\bM(1)$ and $\bMq(1)$.
\par 
By \eqref{paper DT k-domain AE Loop Pressure Equations}, the loop pressure equation is transformed into
\[
    \begin{bmatrix}
        1&-1&1&-1\\ 
    \end{bmatrix}
    \md 
    \begin{bmatrix}
        K_1\\ 
        K_2\\
        K_3\\ 
        K_4
    \end{bmatrix}\cdot
    2\cdot\md
    \begin{bmatrix}
        \M_1(0)\\ 
        \M_2(0)\\
        \M_3(0)\\ 
        \M_4(0)
    \end{bmatrix}
    \begin{bmatrix}
        \M_1(1)\\ 
        \M_2(1)\\
        \M_3(1)\\ 
        \M_4(1)
    \end{bmatrix}
    =\boldsymbol{0},
\]
which is one linear equation about $\bM(1)$.
\par 
By \eqref{paper DT k-domain AE node temperature equations s} and \eqref{paper DT k-domain AE node temperature equations r}, the node temperature equations are transformed into
\[
\begin{aligned}
    &\md
    \begin{bmatrix}
    \Ts_3(0)\\
    \Ts_4(0)
    \end{bmatrix}
    V_\text{3,4}^\text{+}
\begin{bmatrix}
    \M_1(1)\\ 
    \M_2(1)\\
    \M_3(1)\\ 
    \M_4(1)
\end{bmatrix} 
+\md
\begin{bmatrix}
    \Ts_3(1)\\ 
    \Ts_4(1)
\end{bmatrix}
V_\text{3,4}^\text{+}
\begin{bmatrix}
    \M_1(0)\\ 
    \M_2(0)\\
    \M_3(0)\\ 
    \M_4(0)
\end{bmatrix}     
=\\
&V_\text{3,4}^\text{+}
\md
\begin{bmatrix}
    \Touts_1(1)\\ 
    \Touts_2(1)\\
    \Touts_3(1)\\
    \Touts_4(1)
\end{bmatrix}  
\begin{bmatrix}
    \M_1(0)\\ 
    \M_2(0)\\
    \M_3(0)\\ 
    \M_4(0)
\end{bmatrix}  
+
V_\text{3,4}^\text{+}
\md
\begin{bmatrix}
    \Touts_1(0)\\ 
    \Touts_2(0)\\
    \Touts_3(0)\\
    \Touts_4(0)
\end{bmatrix}  
\begin{bmatrix}
    \M_1(1)\\ 
    \M_2(1)\\
    \M_3(1)\\ 
    \M_4(1)
\end{bmatrix},  
\end{aligned}
\]
\[
\begin{aligned}
&\md
\begin{bmatrix}
    \Tr_1(0)\\
    \Tr_2(0)
\end{bmatrix}
    V_\text{1,2}^\text{-}
\begin{bmatrix}
    \M_1(1)\\ 
    \M_2(1)\\
    \M_3(1)\\ 
    \M_4(1)
\end{bmatrix} 
+\md
\begin{bmatrix}
    \Tr_1(1)\\ 
    \Tr_2(1)
\end{bmatrix}
V_\text{1,2}^\text{-}
\begin{bmatrix}
    \M_1(0)\\ 
    \M_2(0)\\
    \M_3(0)\\ 
    \M_4(0)
\end{bmatrix}     
=\\
&
V_\text{1,2}^\text{-}
\md
\begin{bmatrix}
    \Toutr_1(1)\\ 
    \Toutr_2(1)\\
    \Toutr_3(1)\\
    \Toutr_4(1)
\end{bmatrix}  
\begin{bmatrix}
    \M_1(0)\\ 
    \M_2(0)\\
    \M_3(0)\\ 
    \M_4(0)
\end{bmatrix}  
+
V_\text{1,2}^\text{-}
\md
\begin{bmatrix}
    \Toutr_1(0)\\ 
    \Toutr_2(0)\\
    \Toutr_3(0)\\
    \Toutr_4(0)
\end{bmatrix}  
\begin{bmatrix}
    \M_1(1)\\ 
    \M_2(1)\\
    \M_3(1)\\ 
    \M_4(1)
\end{bmatrix}, 
\end{aligned}
\]
which are four linear equations about $\bM(1)$, $\bTs_\text{L}(1)$, $\bTr_\text{S}(1)$. $\bTouts(1)$ and $\bToutr(1)$ have been calculated in \emph{Step 1}.
\par 
By \eqref{paper DT k-domain AE node power equations}, the node power equations are transformed into
\[
    \begin{aligned}
        \begin{bmatrix}
            \Phi_1(1)\\ 
            \Phi_2(1)\\ 
            \Phi_3(1)\\ 
            \Phi_4(1)
        \end{bmatrix}
        &=C_\text{p}\md
        \begin{bmatrix}
            \Mq_1(1)\\ 
            \Mq_2(1)\\
            \Mq_3(1)\\ 
            \Mq_4(1)
        \end{bmatrix}
        \left(
        \begin{bmatrix}
            \Ts_1(0)\\ 
            \Ts_2(0)\\
            \Ts_3(0)\\ 
            \Ts_4(0)
        \end{bmatrix}
        - 
        \begin{bmatrix}
            \Tr_1(0)\\ 
            \Tr_2(0)\\
            \Tr_3(0)\\ 
            \Tr_4(0)
        \end{bmatrix}\right) 
        \\&+C_\text{p}\md
        \begin{bmatrix}
            \Mq_1(0)\\ 
            \Mq_2(0)\\
            \Mq_3(0)\\ 
            \Mq_4(0)
        \end{bmatrix}
        \left(
        \begin{bmatrix}
            \Ts_1(1)\\ 
            \Ts_2(1)\\
            \Ts_3(1)\\ 
            \Ts_4(1)
        \end{bmatrix}
        -
        \begin{bmatrix}
            \Tr_1(1)\\ 
            \Tr_2(1)\\
            \Tr_3(1)\\ 
            \Tr_4(1)
        \end{bmatrix}        
        \right),
    \end{aligned}
\]
which are four linear equations about $\bPhi(1)$, $\bMq(1)$, $\bTs_\text{L}(1)$ and $\bTr_\text{S}(1)$.
\par 
By \eqref{paper DT k domain AE EPS power flow equations1}, \eqref{paper DT k domain AE EPS power flow equations2}, \eqref{paper DT k domain AE EPS power flow equations3}, the power flow equations are transformed into \eqref{paper DT equation k-domain illustrative active power equation}, \eqref{paper DT equation k-domain illustrative reactive power equation} and \eqref{paper DT equation k-domain illustrative voltage equation}, which are seven linear equations about $\bP_\text{PV,R}(1)$, $\bQ_\text{PV,R}(1)$, $\bE_\text{PV,PQ}(1)$ and $\bF_\text{PV,PQ}(1)$.
\par 
By \eqref{paper DT k domain AE Coupling Components1} and \eqref{paper DT k domain AE Coupling Components2}, the coupling component equations are transformed into 
\[
    P_1(1)=-\Phi_1(1)/Z,\quad 
    \Phi_2(1)=c_{\mathrm{m1}}P_3(1),
\] 
which are two linear equations about $P_1(1)$, $\Phi_1(1)$, $\Phi_2(1)$ and $P_3(1)$. $\delta(1)=0$ by definition.
\par 
There are totally twenty-two linear equations. By combining and solving these linear equations, we obtain the first-order DT coefficients of the above mentioned twenty-two variables. Then there remain eight unknown $\bT^{\text{in,s/r}}(1)$ which are the vectors of the DTs of $\btau^\text{in,s/r}$.
\par 
\emph{Step 3:} 
Update $\bT^{\text{in,s/r}}(1)$. Because $\btau^\text{in,s/r}$ equal the temperatures of the nodes that the inlets of the pipes are connected to, we have, for example,
\[
    \Tins_1(1)=\Ts_1(1).
\]Finally, the first-order DT coefficients of all unknown variables are obtained. 
\par 
The above three steps is the $k=1$ case of the more generalized notation
\[
    \begin{aligned}
    &\text{\emph{Step 1:}}\phantom{\ \boldsymbol{A}_0}\bX(k)=\boldsymbol{B}(\bX(0:k-1),\bY(0:k-1),\bZ(0:k-1)),\\
    &\text{\emph{Step 2:}}\ \boldsymbol{A}_0\bY(k)=\boldsymbol{C}(\bX(0:k),\bY(0:k-1),\bW(0:k)),\\ 
    &\text{\emph{Step 3:}}\phantom{\ \boldsymbol{A}_0}\bZ(k)=\boldsymbol{D}(\bY(k)),
    \end{aligned}
\]
where $\bX(k)$ is the vector of $\bTouts(k)$ and $\bToutr(k)$. $\bY(k)$ is the vector of unknown $k$th-order DT coefficients of the variables we solve in \emph{Step 2}. $\bZ(k)$ is the vector of $\bTins(k)$ and $\bTinr(k)$. $\boldsymbol{W}(k)$ is the vector of the $k$th DT coefficients of known variables. $\bX(0:k)$ denotes the combinations of $\bX(0)$ to $\bX(k)$. The same notation also applies to $\bY(k)$, $\bZ(k)$ and $\bW(k)$. We denote by $\bx(t)$, $\by(t)$, $\bz(t)$ and $\bw(t)$ the original variables of these DTs. $\boldsymbol{A}_0$ is the coefficient matrix that relates only to initial values. This follows from the observations in the previous section that the coefficients of the $k$th-order DT coefficients are the zeroth-order DT coefficients. Therefore, inverse of the matrix, $\boldsymbol{A}_0^{-1}$, should be calculated only once in each step. $\boldsymbol{B}(\cdot)$ denotes \eqref{paper DT equation DT of PDE}. $\boldsymbol{C}(\cdot)$ is the known DT coefficients that we move to the right hand side of equations. $\boldsymbol{B}(\cdot)$ and $\boldsymbol{C}(\cdot)$ are mainly comprised of the convolution operations, $\otimes$, of matrices/vectors. $\boldsymbol{D}(\cdot)$ assigns the $k$th order DT coefficients of node temperatures to related $\bT^{\text{in,s/r}}(k)$.
\par 
As illustrated above, we can repeat the above three steps, recursively calculating the DT coefficients from the zeroth-order to the $K$-th order. Finally, the SASs of all variables are obtained.    
\subsection{Adaptive Time Window Control}\label{paper DT section Adaptive Time Window Control}
To strengthen the robustness of the proposed method under big disturbances, an adaptive time window control strategy, which ensures that the temporal step sizes produce results that satisfy the error tolerances, is developed as follows. 
\par 
The idea comes from the philosophy of embedded Runge-Kutta formulas\cite{wanner1993solving}, which constructs two Runge-Kutta formulas with different numerical accuracy while sharing the same function values. Then the difference of these two Runge-Kutta approximations yield an estimate of the local error which can be used for step size control.
\par 
Here, assuming that we have finished the DT-$K$ calculation, we can obtain DT-$K+1$ recursively by one more evaluation of the three steps in Section \ref{paper DT section calculation procedure1}. Then
we can approximate the local truncation error of DT-$K$, which is the Lagrange remainder
\[\frac{(\bx)^{(K+1)}(\xi)}{(K+1)!}(\Delta t)^{K+1},\quad \text{where}\ \xi\in [0,\Delta t],\]
by the difference between DT-$K$ and DT-$K+1$, which is
\[\tilde{\bx}=\bX(K+1)(\Delta t)^{K+1}.\]
\par
To ensure convergence and accuracy requirements, each component of $\tilde{\bx}$ should be within the prescribed error tolerance vector $\varepsilon$, the component of which is defined as
\[\varepsilon_i=\mathrm{Atol}+\min (|x_i(0)|,|x_i(\Delta t)|) \cdot \mathrm{Rtol}\] 
where $\mathrm{Atol}$ and $\mathrm{Rtol}$ are respectively the prescribed absolute and relative error tolerances. {Rtol} is prescribed to control the number of significant figures in the computed
values. $\min(|x_i(0)|,|x_i(\Delta t)|)$ ensures that both ends of the time window $[0,\Delta t]$ satisfy {Rtol}. Atol is prescribed to prevent endless step size diminishing when true values are close to or equal zero.
\par
We take root mean square error
\[err=\sqrt{\frac{1}{n}\sum_{i=1}^n\left(\frac{\tilde{x}_i}{\varepsilon_i}\right)^2 }\] 
to measure the overall relative error of the current time window. Then we compare $err$ with 1. If $err\leq 1$, the current time window is accepted and the computation of next time window is advanced with a new temporal step size $\Delta t_\mathrm{new}$. Else, the current time window is rejected and the computation starts again with $\Delta t_\mathrm{new}$ until $err\leq 1$. 
\par 
There comes the problem of the formulation of $\Delta t_\mathrm{new}$. Based on the temporal accuracy order $K+1$, it is natural to take
\[\Delta t_\mathrm{new}^{\prime}=\Delta t\cdot fac\cdot(1/err)^{1/(K+1)}\]
where $fac$ is a conservative factor smaller than one, which tries to avoid the reject of $\Delta t_\mathrm{new}^{\prime}$. However, it is also observed by numerical experiments that the surge of temporal step sizes increases the probability of the reject of the next time window while the plunge of temporal step sizes to unnecessarily small levels significantly reduces efficiency. Therefore, we should set an upper bound $fac_{\max}$ and a lower bound $fac_{\min}$ to restrain the violent variation of temporal step size, which finally yields
\[\Delta t_\mathrm{new}=\min(fac_{\max}\cdot\Delta t,\max(fac_{\min}\cdot\Delta t,\Delta t_\mathrm{new}^{\prime})).\]
\par
A special case where models containing $\sin$ or $\cos$ pattern should be further discussed here because the even terms of Taylor series of $\sin$ function at $t=k\pi$ ($k\in \mathbb{N}$) equal zero, making $\bX(K)\neq \boldsymbol{0}$ while $\bX(K+1)=\boldsymbol{0}$ if $K$ is odd. The $\cos$ counterpart encounters the same dilemma if $K$ is even. In the cases where $\bX(K+1)=\boldsymbol{0}$, we should perform one more recursive calculation based on DT-$K+1$, and use $\bX(K+2)$ to adjust temporal step sizes. 
\subsection{Overall Pseudocode of Quasi-dynamic Energy Flow Calculation under Disturbances}
We can set disturbances easily by giving the variation of known variables, $\bw(t)$, in Table \ref{paper DT table knowns and unknowns in the illustrative EPS} and \ref{paper DT table knowns and unknowns of DHS nodes}. The variation can be either distinct functions or discrete time-series. In each time window, we first derive $\bW(0):\bW(K)$ and then start the recursive DT calculation. The overall pseudocode of the proposed DT-based quasi-dynamic energy flow calculation method is shown in Algorithm \ref{paper DT algorithm}. 
\begin{algorithm}
    \DontPrintSemicolon
    \SetInd{0.5em}{1em}
    \KwIn{Total simulation time $\mathcal{T}$, $\bx(0)$, $\by(0)$, $\bz(0)$, variation of $\bw(t)$;}
    \KwOut{$\bX(0):\bX(K)$, $\bY(0):\bY(K)$ and $\bZ(0):\bZ(K)$ in each time window;}
    \Begin{
    $t\leftarrow 0,\ \bX(0)\leftarrow\bx(0),\ \bY(0)\leftarrow\by(0),\ \bZ(0)\leftarrow\bz(0)$;\;
    \While{$t<\mathcal{T}$}{
        Update $\boldsymbol{A}_0$, calculate $\boldsymbol{A}_0^{-1}$, and $err\leftarrow 2$;\;
        \While{$err>1$}{
            Derive $\bW(0):\bW(K)$;\;
            \For{$k=1:K+1$}{
                    Perform \emph{Step 1, 2, 3};\;
            }
            Derive $[\tilde{\bx},\tilde{\by}]$ and $[\boldsymbol{\varepsilon}_{\bx},\boldsymbol{\varepsilon}_{\by}]$;\;
            Update $err$;\;
            \If{$err\leq 1$}{
                $t\longleftarrow t+\Delta t$;\; 
            }
            Update $\Delta t_\text{new}$ and $\Delta t\longleftarrow \Delta t_\text{new}$;
        }
    $\bX(0)\leftarrow\sum_{k=0}^K\bX(k)t^k,\ \bY(0)\leftarrow\sum_{k=0}^K\bY(k)t^k,\ \bZ(0)\leftarrow\sum_{k=0}^K\bZ(k)t^k$   
    }
    }
    \caption{DT-based Quasi-dynamic Energy Flow Calculation\label{paper DT algorithm}}
\end{algorithm}
\subsection{Some special cases}
\subsubsection{Compound node types}
\begin{figure}[!h]
    \centering
    \includegraphics[width=2.7in]{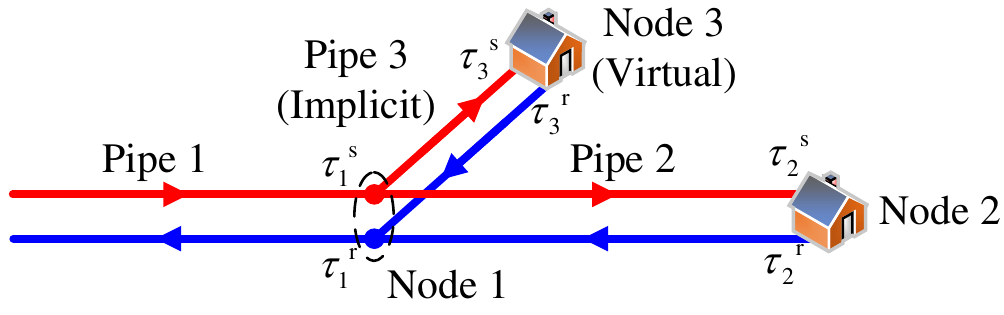}
    \caption{Dealing with compound node types.}\label{paper DT figure compound node}
\end{figure}

In, for example, the Barry Island system\cite{Liuxuezhi2016}, some loads are located in intermediate nodes. Thereby, according to Table \ref{paper DT table knowns and unknowns}, $\taur$ of these nodes are both known and unknown, which is contradictory. Actually, the real case is, load nodes are connected to intermediate nodes through \emph{implicit pipes} whose length is \SI{0}{\metre}. As shown in Fig. \ref{paper DT figure compound node}, we can extract a virtual load node and then the original compound node becomes an intermediate one.
\par    
Since the length of Pipe 3 is \SI{0}{\metre}, we can not model Pipe 3 by PDE and hence, $\Touts_3(k+1)$ and $\Toutr_3(k+1)$ can not be obtained in \emph{Step 1} of the SAS-derivation framework.
\par
Instead, we have two additional equations,
\[\tauouts_3=\taus_1,\ \tauoutr_3=\taur_3.\]
After performing DT, we have
\[\Touts_3(k+1)=\Ts_1(k+1),\ \Toutr_3(k+1)=\Tr_3(k+1).\]
There are two extra variables and two extra equations. So we still can successfully perform \emph{Step 2} of the SAS-derivation framework by adding the above two equations to $\boldsymbol{C}(\cdot)$.    
\par 
The cases where source nodes are located in intermediate nodes can be dealt with likewise.
\subsubsection{Reverse mass flow rates}
DHSs are directed networks, that is, if mass flow rate of some pipe is reversed at some time during the simulation, then the mass flow continuity equations, loop pressure equations and node temperature equations will have different formulae. Fortunately, these equations are depicted by node/loop incidence matrices in this paper. We can obtain the post-reverse equations by reversing columns related to these reverse pipes in those matrices. As for the semi-discrete PDE, we should exchange the inlet and outlet in Fig. \ref{paper DT figure semi-discrete}, and reverse the computation sequence.
\par
Having finished the calculation in one time window, we judge if, at the end of the current time window $[t_0, t_1]$, some $\m_i<0$. If there is, then we should find the zero $t'$ of univariate polynomial equation
\[\sum_{k=0}^KM_i(k)t^k=0,\]
which can be solved by the Newton-Raphson or bisection method. This is because, after $t>t'$, the equations differ from the original ones.
\par 
These special cases were considered in the following Barry Island case study.
\section{Case Studies}\label{paper DT section case studies}
In this section, we tested the proposed method on 1) a real DHS with measured data, 2) the Barry Island system and 3) a 225-node-118-bus big system. In the first system, we showed that the proposed method can effectively reduce dissipative and dispersive errors, which improves the accuracy of PDE solutions. In the second system, we compared the proposed method with iteration ones in terms of accuracy and time performance. We analyzed the reason of efficiency improvement of the proposed method in details and studied the impact of tuning parameters of the proposed method on computational performance. In the last system, we mainly tested the robustness and convergence of the proposed method under severe disturbances. All the tests were programmed with MATLAB R2022a on a desktop computer equipped with AMD Ryzen 7 3700x CPU and 64GB RAM.
\subsection{A Real DHS with Measured Data}
\label{paper DT section A Real DHS with Measured Data}
\begin{figure}[!h]
    \centering
    \includegraphics[width=3.2in]{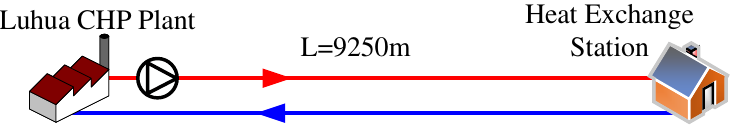}
    \caption{A real DHS located in Shijiazhuang, Hebei Province, China.}\label{paper DT figure single pipe}
\end{figure}
\par As shown in Fig. \ref{paper DT figure single pipe}, the real DHS, located in the suburb of Shijiazhuang, Hebei Province, China, consists of a pair of supply and return pipes that connect the CHP plant and the heat exchange station. The parameters and measured data are given in\cite{paperDT_supplementary_material}. 
The spatial and temporal step sizes were set to be 370 meters and 180 seconds, respectively. IU scheme, SOE scheme, node method (NM) and DT-20 were performed.
\par 
We first fixed $\dot{m}$ to be \SI{2543.5}{\kilo\gram\per\second} and gave step boundary condition. The temperature of the inlet, denoted by $\tau_1$, increased from \SI{90.1725}{\celsius} to \SI{92.0000}{\celsius} at $t=$\SI{10}{\hour}. The exact solution of \eqref{paper DT equation PDE heat pipe model}, denoted by REF below, was obtained by the characteristic line method, which is
\begin{equation}
    \tau(x,t)=(1-e^{-\frac{\lambda}{C_\mathrm{p}\dot{m}}x})\tau^\mathrm{amb}+e^{-\frac{\lambda}{C_\mathrm{p}\dot{m}}x}\tau(0,t-\frac{\gamma\rho}{\dot{m}}x).
\end{equation}
\par
\begin{figure}[h]
    \centering
    \includegraphics[width=3.3in]{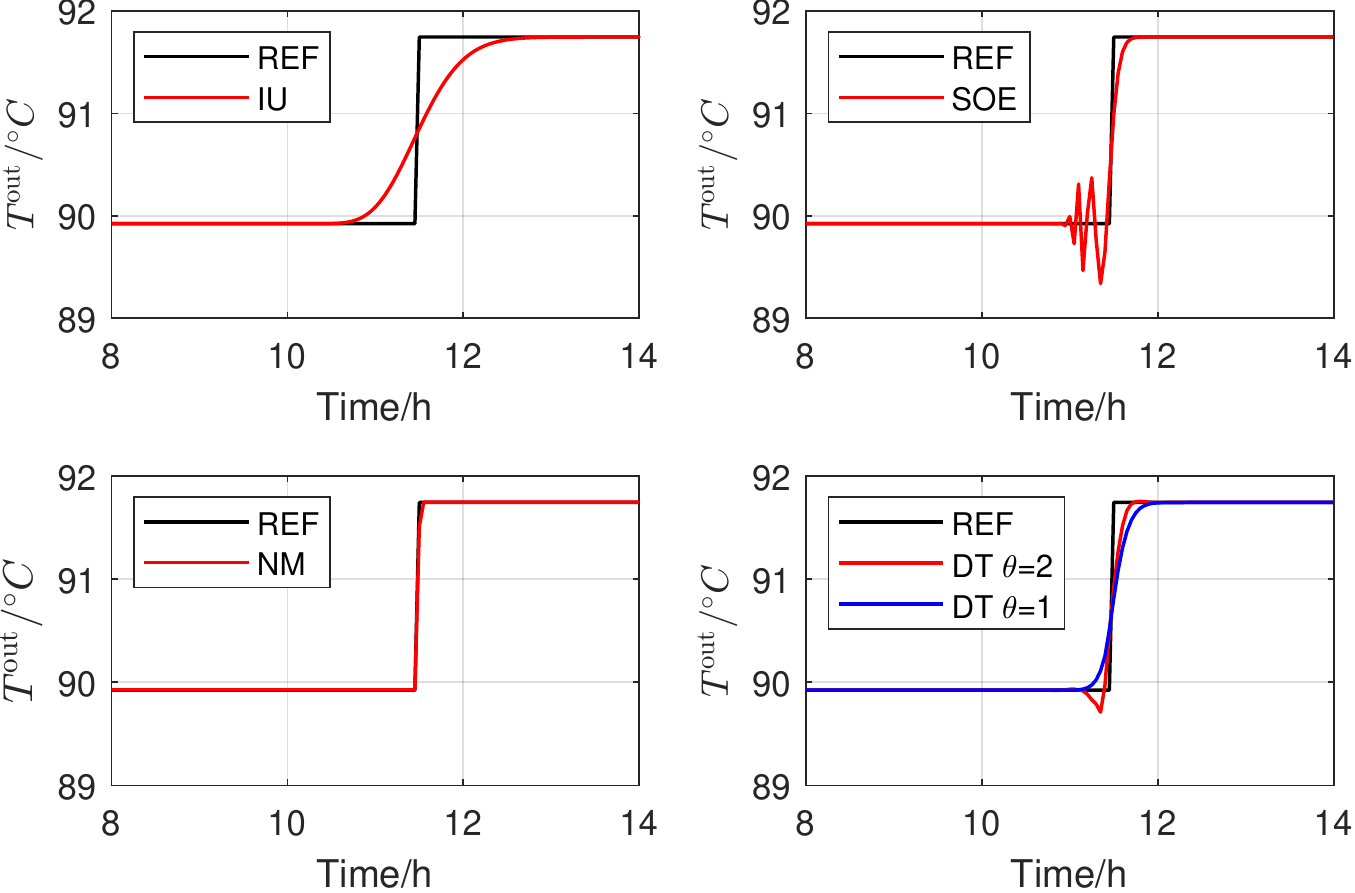}
    \caption{Accuracy comparison on step function.}\label{paper DT figure Method comparison of step function}
\end{figure}

 
As shown in Fig. \ref{paper DT figure Method comparison of step function}, solution by the IU scheme was severely smeared because of its big dissipative errors. Obviously, high frequency components in the solution were damped out. Meanwhile, serious oscillation ruined the solution by the SOE scheme due to its dispersive errors. The solution by DT with $\theta=2$ had small dissipative errors while it revealed small dispersive error in front of the rising edge. The solution by DT with $\theta=1$ had relatively big dissipative error but no dispersive error. The node method performed best in the test, showing no dissipative and dispersive errors.
\begin{figure}
    \centering
    \subfloat{\includegraphics[width=3.2in]{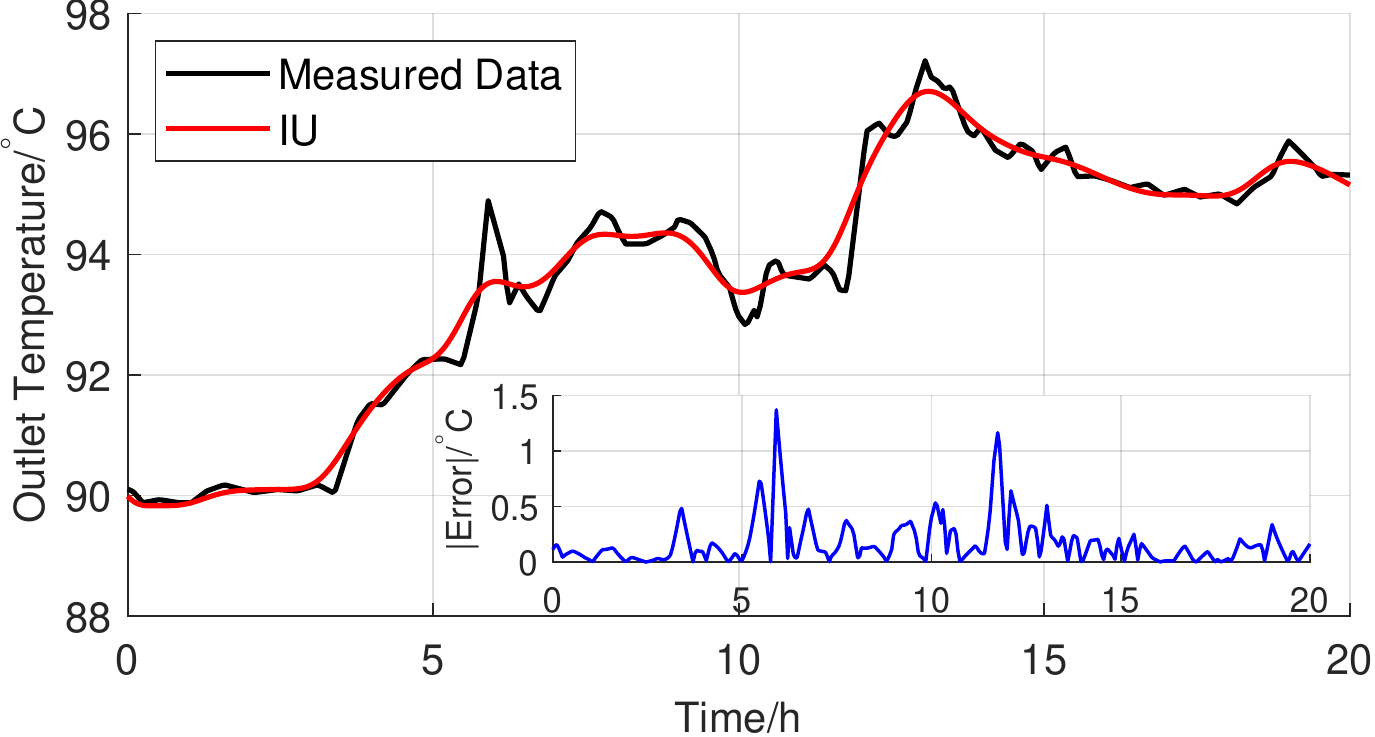}}\\
    \subfloat{\includegraphics[width=3.2in]{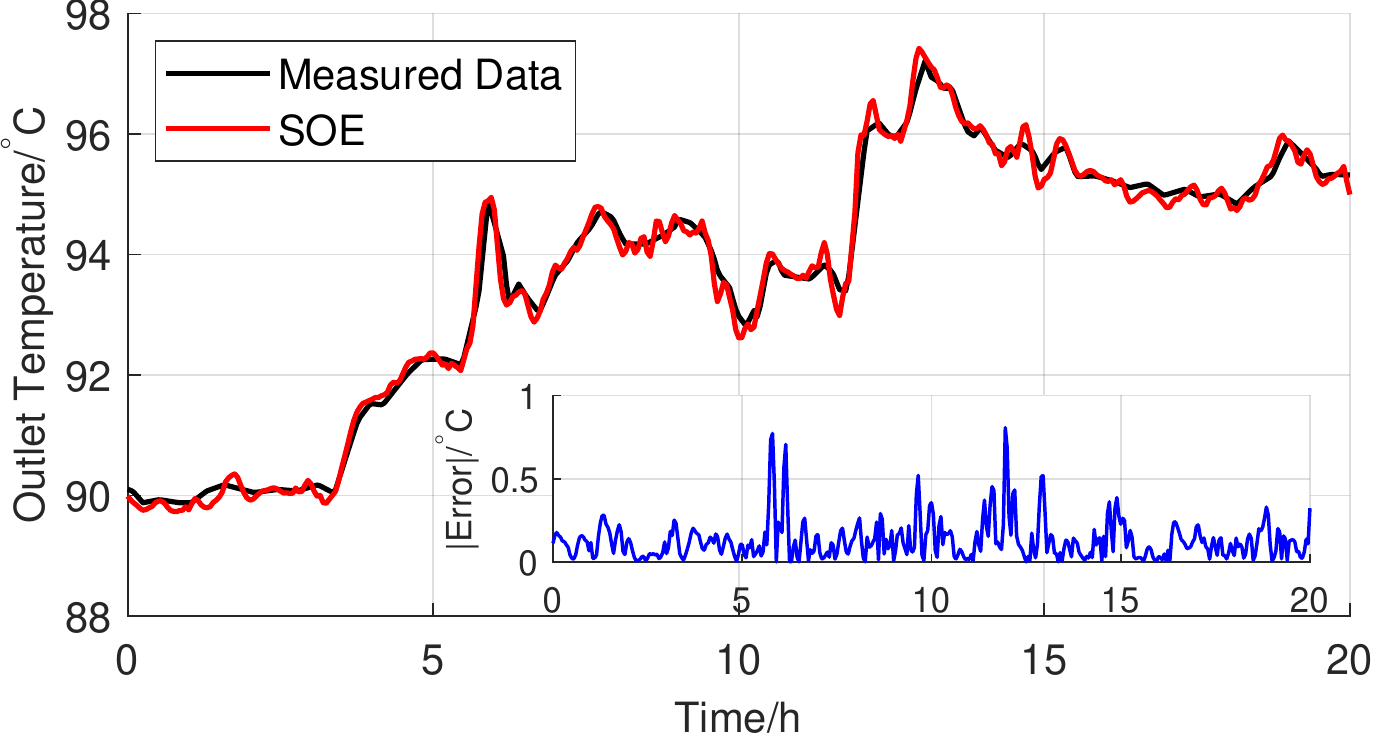}}\\
    \subfloat{\includegraphics[width=3.2in]{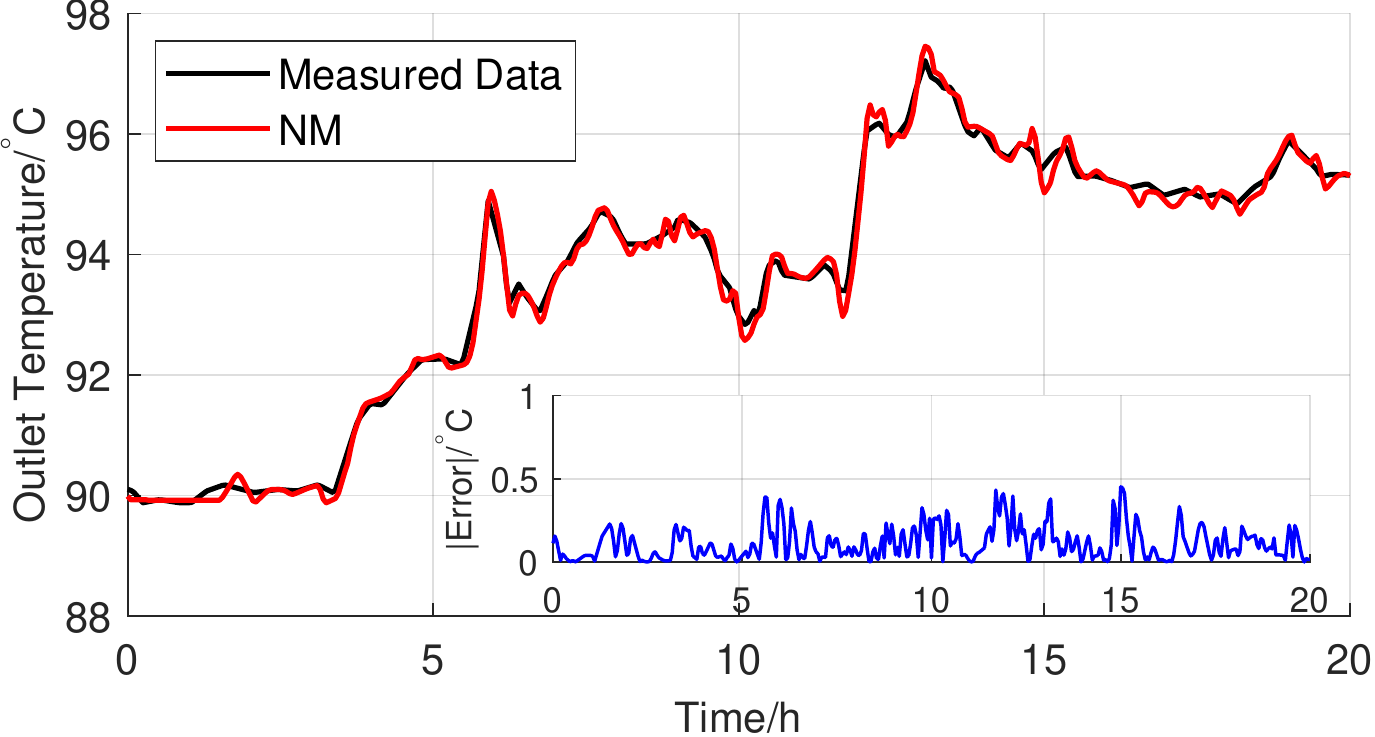}}\\
    \subfloat{\includegraphics[width=3.2in]{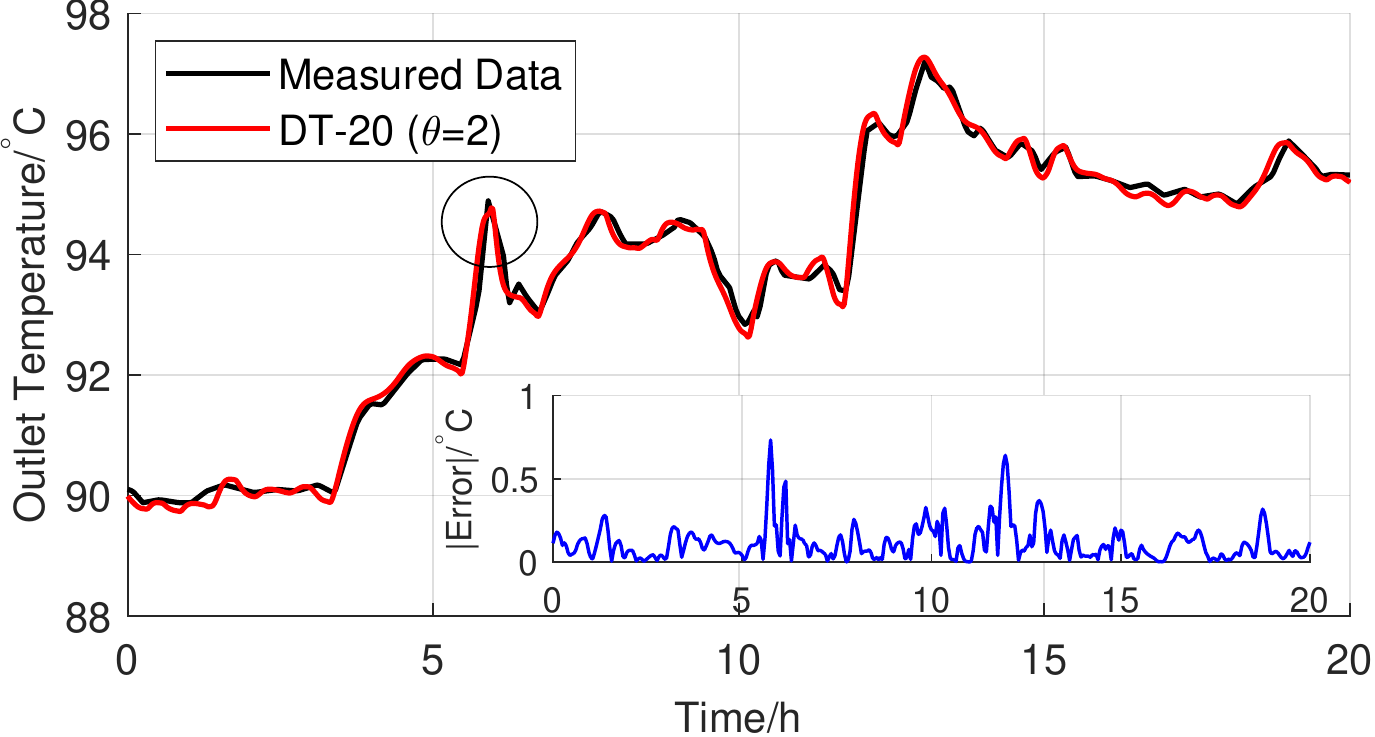}}\\   \subfloat{\includegraphics[width=3.2in]{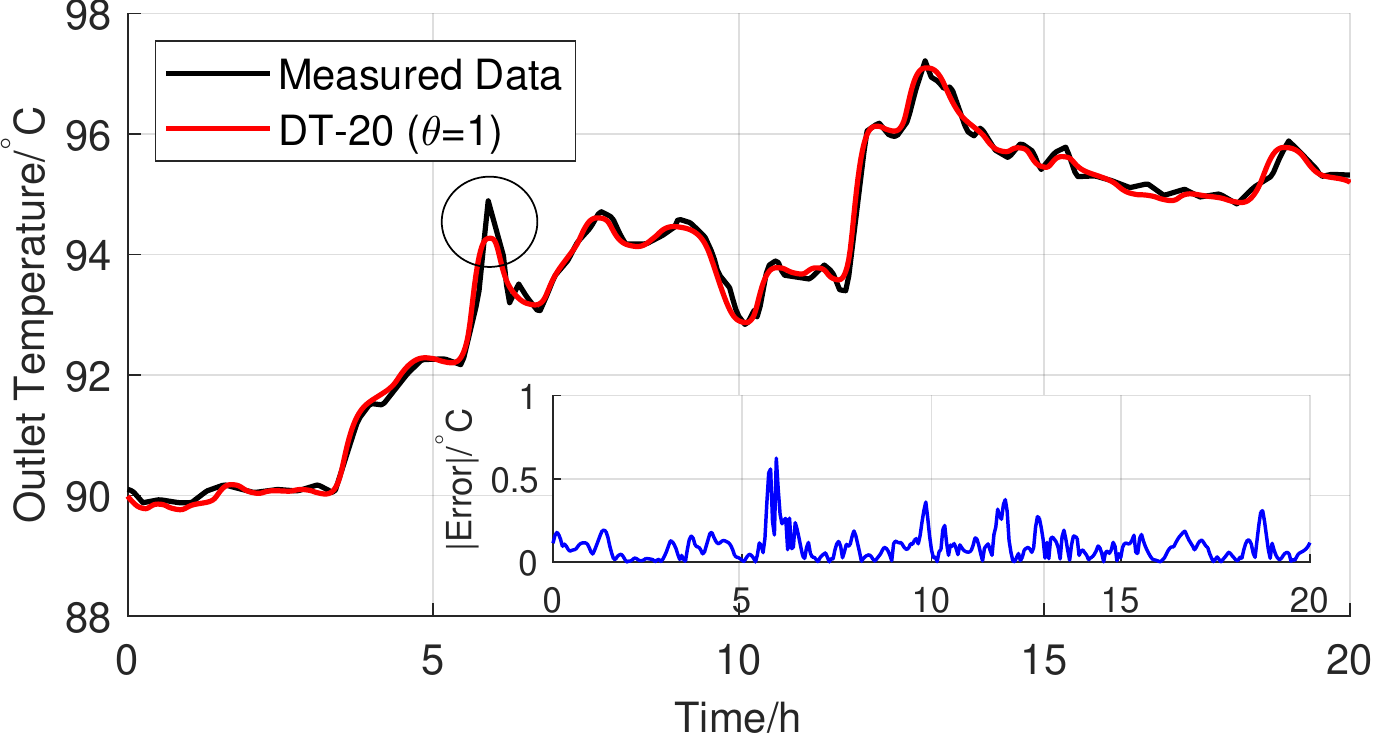}}\\
    \caption{Accuracy comparison on a real DHS.}\label{paper DT figure Method comparison in a practical DHN}
\end{figure}
\par 
Next, we varied both $\dot{m}$ and $\tau_1$. The results were compared with the measured data in Fig. \ref{paper DT figure Method comparison in a practical DHN}. We first knocked out the IU scheme, because it is of first-order accuracy only and displayed big dissipative errors, omitting most of the details of temperature variation. Solutions by the SOE scheme and NM portrayed and tracked the temperature variations well, but small oscillations could not be avoided. On the contrary, solutions by DT got rid of these fake oscillations effectively. Compared with $\theta=2$, $\theta=1$ produced more conservative results. As circled in Fig. \ref{paper DT figure Method comparison in a practical DHN}, the real oscillation was damped with $\theta=1$. However, the root mean square errors (RMSEs) between the solutions and the measured data, which are computed in Table \ref{paper DT table case1 RMSE}, showed that DT with $\theta=1$ was the most accurate and was trailed by the NM and DT with $\theta=2$. The exclusive TVD property granted the proposed method accuracy superiority over others and we think that DT with $\theta=1$ should be adopted in this case.
\begin{table}[!h]
    \centering
    \caption{RMSEs against Measured Data}
    \begin{tabular}{ccccc}
        \hline
        IU&SOE&NM&DT($\theta=2$)&DT($\theta=1$)\\
        \hline
        0.2607&0.1773&0.1484&0.1546&0.1253\\
        \hline
    \end{tabular}\label{paper DT table case1 RMSE}
\end{table}
\subsection{The Barry Island System}
\begin{figure*}
    \centering
    \includegraphics[width=7in]{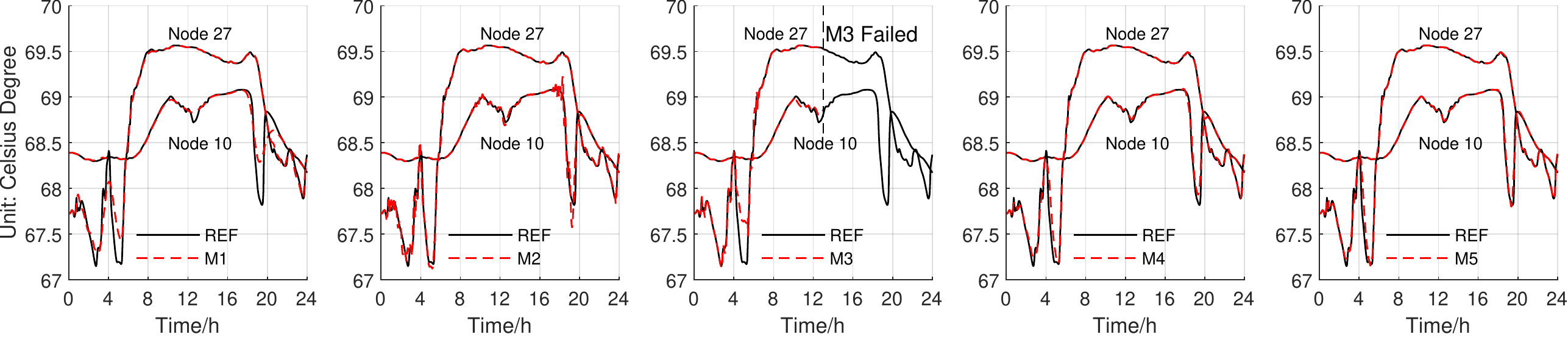}
    \caption{Node supply temperature in Barry  Island case.}
    \label{paper DT figure Node supply temperature}
\end{figure*}
\begin{figure*}
    \centering
    \subfloat{\includegraphics[width=7in]{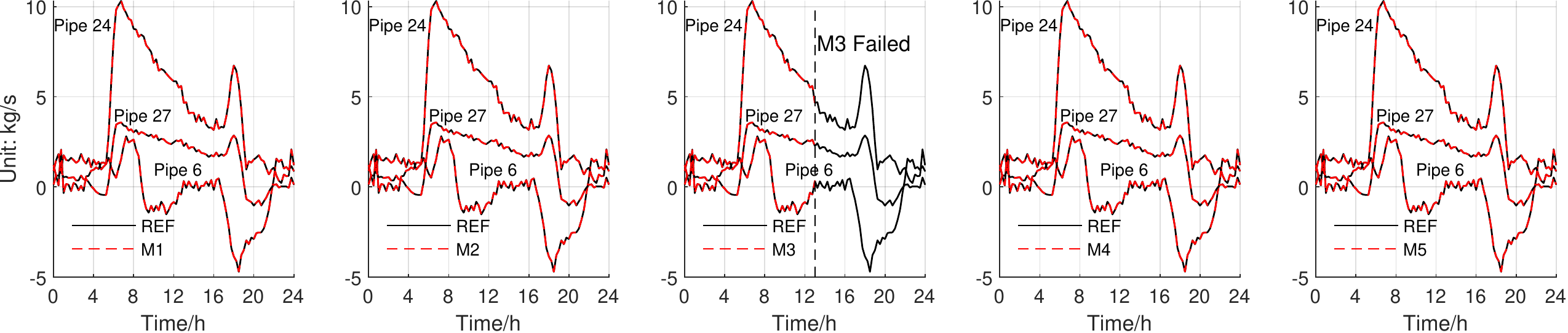}}
    \caption{Mass flow rate in Barry  Island case.}
    \label{paper DT figure mass flow rate}
\end{figure*}
\begin{figure*}
    \centering
    \subfloat{\includegraphics[width=7in]{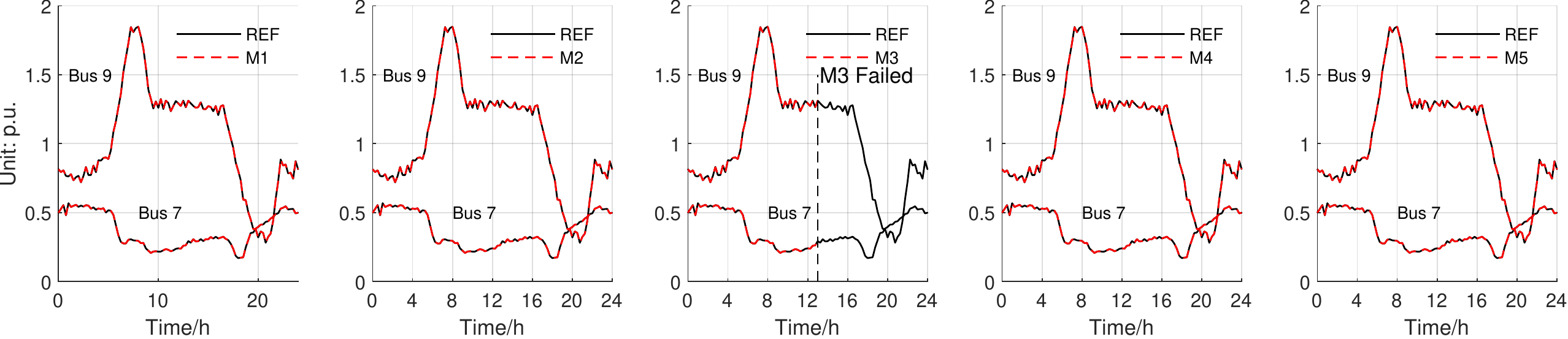}}
    \caption{Electric Power in Barry  Island case.}
    \label{paper DT figure pe}
\end{figure*}
\par
To test the accuracy, efficiency and convergence performance of the proposed method, the famous Barry Island system was used here. Each heating and electric load was given a typical 24-hour variation trajectory, with \SI{15}{\minute} as the interval. These load trajectories are classified into four types, including Supermarkets, Guaranteed Savings Buildings, Hotels and Industrial Factories. The reader can refer to \cite{qinCombinedElectricHeat2021} for design philosophy of the test bed and \cite{paperDT_supplementary_material} for detailed parameters.
\par
The following methods were performed.
\begin{itemize}
    \item M1---The HE-FBI iteration solver proposed in \cite{Qin2019}, which solves PDE with IU scheme.
    \item M2---Replace the IU scheme in M1 by SOE scheme.
    \item M3---Replace the IU scheme in M1 by NM.
    \item M4---DT-6 with $\theta=1$.
    \item M5---DT-6 with $\theta=2$.
\end{itemize}
\par
The spatial step size for M1, M2, M4 and M5 was \SI{20}{\meter}. The temporal step size for M1, M2 was \SI{60}{\second}. The convergence performance of M3 is unsatisfactory. For M3, only temporal step size \SI{10}{\second} produced relatively good results, which was adopted here. M4 and M5 were performed with adaptive windows, but results with fixed time windows were obtained from SASs with temporal step size \SI{60}{\second}. Error tolerance was set to be 1e-9. 
\par
The Reference method (REF) generated reference values with spatial and temporal step sizes \SI{2}{\meter} and \SI{3}{\second} respectively. The PDE \eqref{paper DT equation PDE heat pipe model} is first spatially discretized by \eqref{paper DT equation time domain TVD scheme} and then temporally discretized by the Dormand-Prince5 Runge-Kutta formula \cite{DOPRI45}. \cite{AlexanderKurganov2000} has verified the accuracy of the PDE solver. The nonlinear algebraic equations and the discretized PDEs were alternatively solved as \cite{Qin2019} does in each of the seven stages of Dormand-Prince5.
\par
The EPS power flow calculations in M1, M2, M3 and REF were performed with the state-of-the-art Matpower V7.1\cite{matpowerACpowerflow}.
\par 
The simulation results of $\bTs$, $\bm$ and $\bp$ are shown in Fig. \ref{paper DT figure Node supply temperature}-\ref{paper DT figure pe}. The dissipative errors of M1 and the dispersive errors of M2 were still obvious in temperatures. Different from the preceding fixed direction case, M3 could not accurately depict the temperature variations in the limited convergent range. However, M4 and M5 obtained fairly good temperature results. Though mass flow rates were reversed several times in this case, M1, M2, M4 and M5 still produced accurate $\bm$ and $\bp$ trajectories.
\par 
The maximum RMSEs of each variable were listed in Table \ref{paper DT table Barry RMSE}. M5 had the smallest errors, which was trailed by M4. On one hand, the proposed method diminished dissipative and dispersive errors effectively with the TVD property. On the other, the high order temporal accuracy, which was 20 here, and the strict time window control strategy impeded the growth of errors. These facts explain the accuracy performance of the proposed method from the theoretical perspective.
\begin{table}[h]
    \centering
    \caption{Maximum RMSEs}
    \begin{tabular}{c|ccccc}
    \hline
    Variable& M1         &     M2     & M3 & M4 & M5\\ 
    \hline
    $\bTs$ (\si[]{\celsius})&1.01e-1&7.57e-2&\textbackslash&5.64e-2&\cellcolor{lightgray}5.21e-2\\
    $\bTr$ (\si[]{\celsius})&3.41e-2&3.52e-2&\textbackslash&2.13e-2&\cellcolor{lightgray}1.80e-2\\
    $\bm$ (\si[]{\kilo\gram\per\second})&7.27e-3&4.99e-3&\textbackslash&3.82e-3&\cellcolor{lightgray}3.17e-3\\
    $\bphi$ (\si[]{\watt})&1.60e+3&1.04e+3&\textbackslash&7.49e+2&\cellcolor{lightgray}5.96e+2\\
    $\bp$ (p.u.)&1.98e-4&1.28e-4&\textbackslash&9.30e-5&\cellcolor{lightgray}7.39e-5\\
    $\bee$ (p.u.)&1.34e-7&8.45e-8&\textbackslash&6.13e-8&\cellcolor{lightgray}4.89e-8\\
    $\bff$ (p.u.)&2.87e-6&1.86e-6&\textbackslash&1.35e-6&\cellcolor{lightgray}1.07e-6\\ 
    \hline
    \end{tabular}    \label{paper DT table Barry RMSE}
\end{table}
\begin{table}[h]
    \centering
    \begin{threeparttable}
    \caption{Time Performance}
    \begin{tabular}{c|ccccc}
    \hline
    Time Cost (\si[]{\second})&M1&M2&M3&M4&M5\\ 
    \hline
    Total&36.17&37.64&\textbackslash&27.30&\cellcolor{lightgray}20.86\\
    Per Step&0.025&0.026&\textbackslash&0.019&\cellcolor{lightgray}0.018\\
    \hline
    \end{tabular}\label{paper DT table Barry Time Performance}
    \begin{tablenotes}
        \item[1] M1 and M2 took 1440 steps, M4 took 1452 steps, M5 took 1183 steps.
    \end{tablenotes}        
    \end{threeparttable}
\end{table}
\begin{table}[h]
    \centering
    \begin{threeparttable}
    \caption{Components of Per-Step Time Cost (\si[]{\second})}
    \begin{tabular}{c|cc}
    \hline
    Simulation Routines&M1&M5\\ 
    \hline
    Matrix Update and Factorization&0.0171&\cellcolor{lightgray}0.0011\\
    Arithmetic Operation\tnote{1}&\cellcolor{lightgray}0.0068&0.0128\\
    Others&\cellcolor{lightgray}0.0007&0.0037\\
    \hline
    \end{tabular}\label{paper DT table Barry Time Cost per Step}
    \begin{tablenotes}
        \item[1] For M1, it mainly contains the recursive calculation of pipe temperatures. For M5, it mainly contains the convolution operation, $\otimes$, of matrices/vectors.
    \end{tablenotes}        
    \end{threeparttable}
\end{table}
\begin{table}
    \centering
    \begin{threeparttable}
    \caption{Average Number of Iterations of M1}
    \begin{tabular}{c|c}
    \hline
    Iteration Loops&Number of iterations\\ 
    \hline
    IES loop&10.60\\
    DHS loop&2.24\\
    Hydraulic loop&1.92\\
    \hline
    \end{tabular}\label{paper DT table Barry Average Number of Iterations} 
    \end{threeparttable}
\end{table}
\begin{figure*}
    \centering
    \subfloat{\includegraphics[width=2.38in]{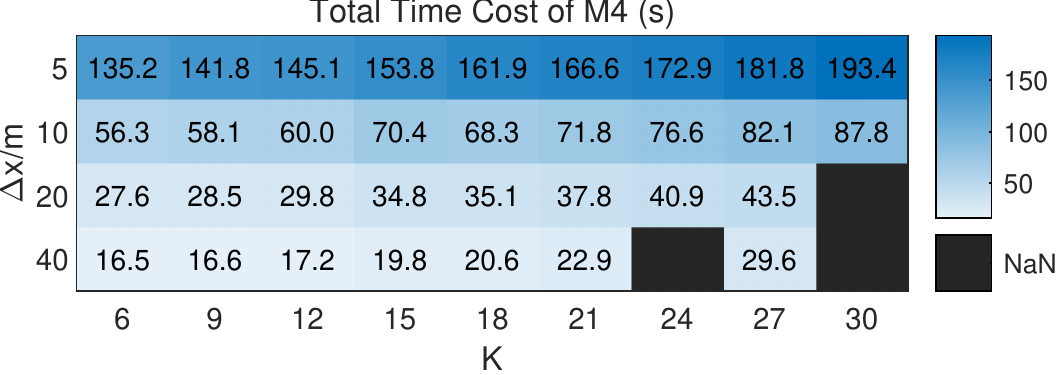}}
    \subfloat{\includegraphics[width=2.38in]{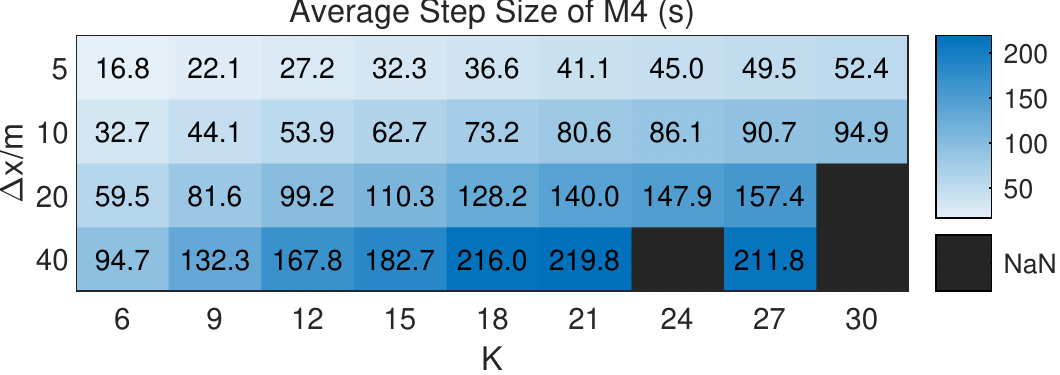}}
    \subfloat{\includegraphics[width=2.38in]{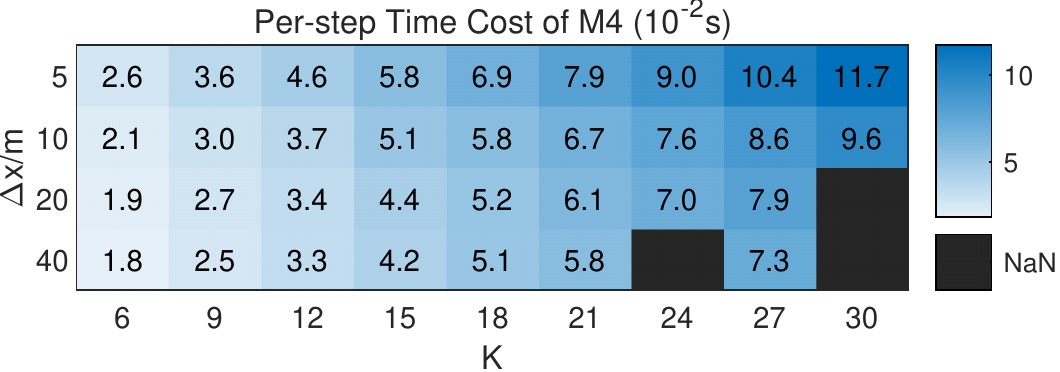}}\\
    \subfloat{\includegraphics[width=2.38in]{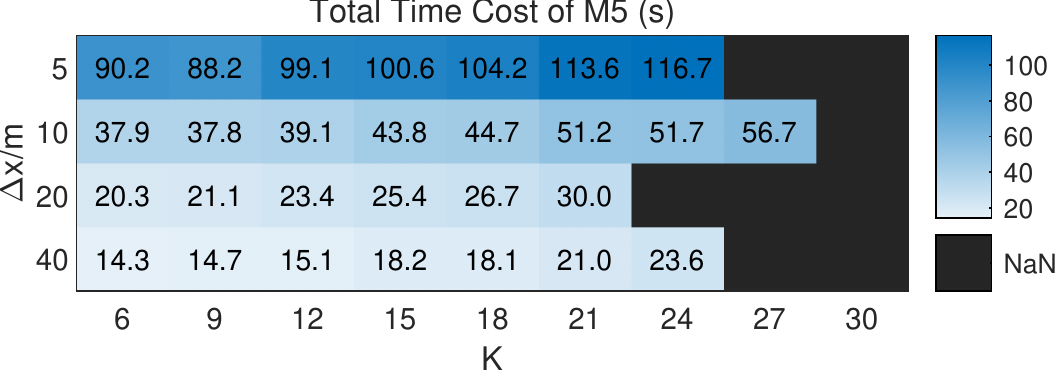}}
    \subfloat{\includegraphics[width=2.38in]{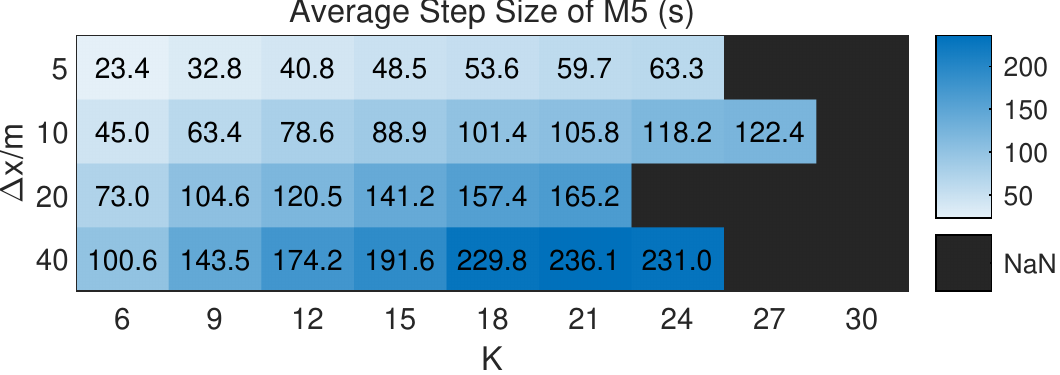}}
    \subfloat{\includegraphics[width=2.38in]{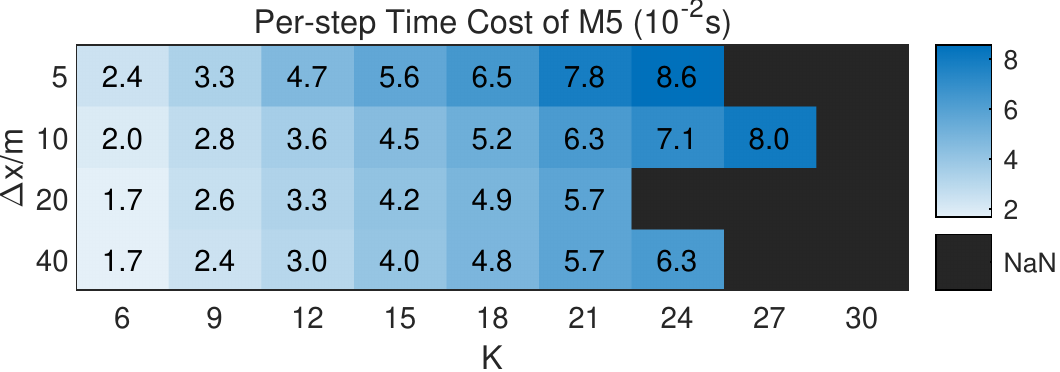}}
    \caption{Computational performance in terms of spatial step size and $K$.}
    \label{paper DT figure Computational performance in terms of h and K}
\end{figure*}
\par 
The time performance of M1-M5 were compared in Table \ref{paper DT table Barry Time Performance}. M5 spent the shortest time finishing the simulation, trailed first by M4 and then M1, M2. 
The efficiency improvement not only came from the adaptive time window strategy, i.e. fewer total steps, but also from the smaller per-step time cost. To explain this, the components of time cost in each step were profiled and listed in Table \ref{paper DT table Barry Time Cost per Step}, with M1 and M5 serving as the representative of iterative and non-iterative solvers respectively. The update and factorization of Jacobian/coefficient matrices occupy the most computation overhead of M1 because, as shown in Table \ref{paper DT table Barry Average Number of Iterations}, tens of iterations are required in each step of M1. Different from M1, the time cost of M5 in matrix computation was tiny. This followed from the fact, which has been pointed out in Section \ref{paper DT section calculation procedure1}, that matrix update and factorization should be performed only once in each step of M5. Although it took M5 much more time to perform the arithmetic operations, which mainly consisted of $K$ times convolution operations of matrices/vectors, the overall computational burden was still small. Therefore, the proposed method (M4) was able to perform more efficiently when the spatial and temporal step sizes resembled M1 and M2's. With the adaptive window strategy, the proposed method (M5) could further improve the overall efficiency.
\par  Even if we varied different $\Delta x$, the proposed method still had better per-step time performance as shown in Table \ref{paper DT table Barry Per Step Time Cost under Different h}. We believe that $K=6$ has the best time performance in this system. Because it was observed from Fig. \ref{paper DT figure Computational performance in terms of h and K} that bigger $K$ meant bigger average step size, but the smaller number of total steps was offset by the spiking per-step time cost, finally decreasing the overall efficiency. It should be also noted that infinitely enlarging $K$ led to divergence instead of infinite temporal step size. This is because M4 and M5 are conditionally stable and hence, have limited convergence region. Their convergence region, which was reflected by the average temporal step size $\overline{\Delta t}$, depended on the choice of $\Delta x$, and we found that bigger $\Delta x$ allowed bigger $\overline{\Delta t}$. Therefore, the adaptive time window strategy is critical to ensure computation convergence when we do not know what $\Delta t$ ensures convergence. The choice of $\theta$ also influences the computation performance. M4 and M5 had similar per-step time cost, but M5 admitted bigger $\overline{\Delta t}$, which achieved higher overall efficiency in this system. 
\begin{table}
    \centering
    \caption{Per-step Time Costs under Different Spatial Step Size (\SI{e-2}{\second})}
    \begin{tabular}{ccccc}
    \hline
    &\SI{5}{\metre}&\SI{10}{\metre}&\SI{20}{\metre}&\SI{40}{\metre}\\ 
    \hline
    M1&3.9&2.9&2.4&2.2\\ 
    M2&4.0&3.0&2.5&2.2\\ 
    M4&2.6&2.1&1.9&1.8\\
    M5&\cellcolor{lightgray}2.4&\cellcolor{lightgray}2.0&\cellcolor{lightgray}1.7&\cellcolor{lightgray}1.7\\
    \hline
    \end{tabular}\label{paper DT table Barry Per Step Time Cost under Different h}       
\end{table}
\begin{table}[]
    \centering
    \caption{Maximum RMSEs in Scenario 1}
    \begin{tabular}{c|ccccc}
    \hline
    Variable& M1         &     M2     & M3 & M4 & M5\\ 
    \hline
    $\bTs$ (\si[]{\celsius})&1.22e-1&7.69e-2&\cellcolor{lightgray}2.76e-2&4.56e-2&\textbackslash\\
    $\bTr$ (\si[]{\celsius})&1.96e-4&1.50e-4&5.59e-3&\cellcolor{lightgray}6.34e-5&\textbackslash\\
    $\bm$ (\si[]{\kilo\gram\per\second})&1.96e-1&8.65e-2&7.00e-2&\cellcolor{lightgray}4.50e-2&\textbackslash\\
    $\bphi$ (\si[]{\watt})&1.71e+4&7.56e+3&1.17e+4&\cellcolor{lightgray}4.95e+3&\textbackslash\\
    $\bp$ (p.u.)&1.54e-2&6.80e-3&1.05e-2&\cellcolor{lightgray}4.45e-3&\textbackslash\\
    $\bee$ (p.u.)&5.35e-6&2.36e-6&3.64e-6&\cellcolor{lightgray}1.54e-6&\textbackslash\\
    $\bff$ (p.u.)&6.30e-5&2.78e-5&4.29e-5&\cellcolor{lightgray}1.82e-5&\textbackslash\\ 
    \hline
    \end{tabular}    \label{paper DT table Big RMSE1}
\end{table}
\begin{table}[]
    \centering
    \caption{Maximum RMSEs in Scenario 2}
    \begin{tabular}{c|ccccc}
    \hline
    Variable& M1         &     M2     & M3 & M4 & M5\\ 
    \hline
    $\bTs$ (\si[]{\celsius})&2.32e-2&3.32e-3&1.38e-2&5.55e-4&\cellcolor{lightgray}2.62e-4\\
    $\bTr$ (\si[]{\celsius})&1.95e-4&5.95e-5&5.59e-3&1.25e-5&\cellcolor{lightgray}8.55e-6\\
    $\bm$ (\si[]{\kilo\gram\per\second})&3.92e-3&6.56e-4&5.51e-2&1.66e-4&\cellcolor{lightgray}8.67e-5\\
    $\bphi$ (\si[]{\watt})&5.10e+2&8.69e+1&8.99e+3&1.97e+1&\cellcolor{lightgray}1.22e+1\\
    $\bp$ (p.u.)&4.59e-4&7.82e-5&8.10e-3&1.77e-5&\cellcolor{lightgray}1.10e-5\\
    $\bee$ (p.u.)&1.86e-7&3.18e-8&3.18e-6&6.69e-9&\cellcolor{lightgray}4.28e-9\\
    $\bff$ (p.u.)&1.85e-6&3.14e-7&3.27e-5&7.23e-8&\cellcolor{lightgray}4.44e-8\\ 
    \hline
    \end{tabular}    \label{paper DT table Big RMSE2}
\end{table}
\begin{table}
    \centering
    \caption{Time Performance}
    \begin{tabular}{c|ccccc}
    \hline
    Time Cost (\si[]{\second})&M1&M2&M3&M4&M5\\ 
    \hline
    Scenario 1&180.35&179.54&402.15&\cellcolor{lightgray}146.40&\textbackslash\\
    Scenario 2&264.57&252.04&848.18&216.24&\cellcolor{lightgray}150.18\\
    \hline
    \end{tabular}\label{paper DT table Big Time Performance} 
\end{table}
\subsection{A 225-DHS-Node and 118-EPS-Bus HE-IES}\label{paper DT case study B}
The 225-DHS-Node and 118-EPS-Bus HE-IES was constructed by integrating a real DHS from \cite{Gong2019} and an EPS from \cite{zhangdong2006}. The original DHS was modified by adding a loop. An extraction steam turbine acts as the slack node 1 of DHS and the PV bus 118 of EPS. A gas turbine acts as the source node 224 of DHS and the slack bus 1 of EPS. The readers can refer to \cite{paperDT_supplementary_material} for detailed parameters. 
\subsubsection{Accuracy and efficiency}
M1-M2 with $\Delta x=$\SI{25}{\metre} and $\Delta t=$\SI{60}{\second}, M2 with $\Delta t=$\SI{60}{\second}, M4-M5 with $K=20$ and $\Delta x=$\SI{25}{\metre}, and REF with $\Delta x=$\SI{5}{\metre} and $\Delta t=$\SI{2}{\second} were performed in the following two scenarios. 
\begin{itemize}
    \item Scenario 1: $\mathcal{T}=$\SI{4}{\hour}. $\Ts_1$ increased from \SI{85}{\celsius} at $t=$\SI{600}{\second} to \SI{86}{\celsius} at $t=$\SI{660}{\second}.
    \item Scenario 2: $\mathcal{T}=$\SI{4}{\hour}. Active load of PQ bus 63 to 77 and PQ bus 100 to 111 changed sinusoidally with amplitude half of its original value and period $0.35\mathcal{T}$. The disturbance started at $t=\mathcal{T}/40$ and ended at $t=36\mathcal{T}/40$.
\end{itemize}    
\par 
Accuracy and time performance of M1-M5 were compared in Table \ref{paper DT table Big RMSE1} to Table \ref{paper DT table Big Time Performance}. M5 failed in scenario 1, so we think that M5 is not suitable for cases where there are steep temperature changes. M3 obtained the most accurate $\bTs$ results since, as shown in Fig. \ref{paper DT figure Method comparison of step function}, it can effectively get rid of dissipative and dispersive errors in this case. Except for $\bTs$, the proposed algorithm obtained more accurate results with smaller total time costs.
\begin{figure}[!h]
    \centering
    \includegraphics[width=3.4in]{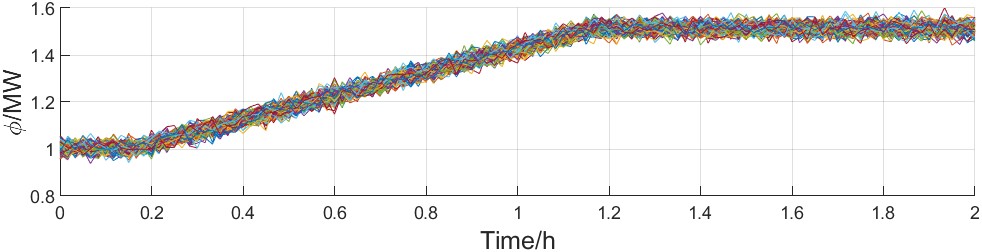}
    \caption{Load curve when loading level=150\%.}\label{paper DT figure robustness_phi}
\end{figure}
\begin{figure}[!h]
    \centering
    \includegraphics[width=3.4in]{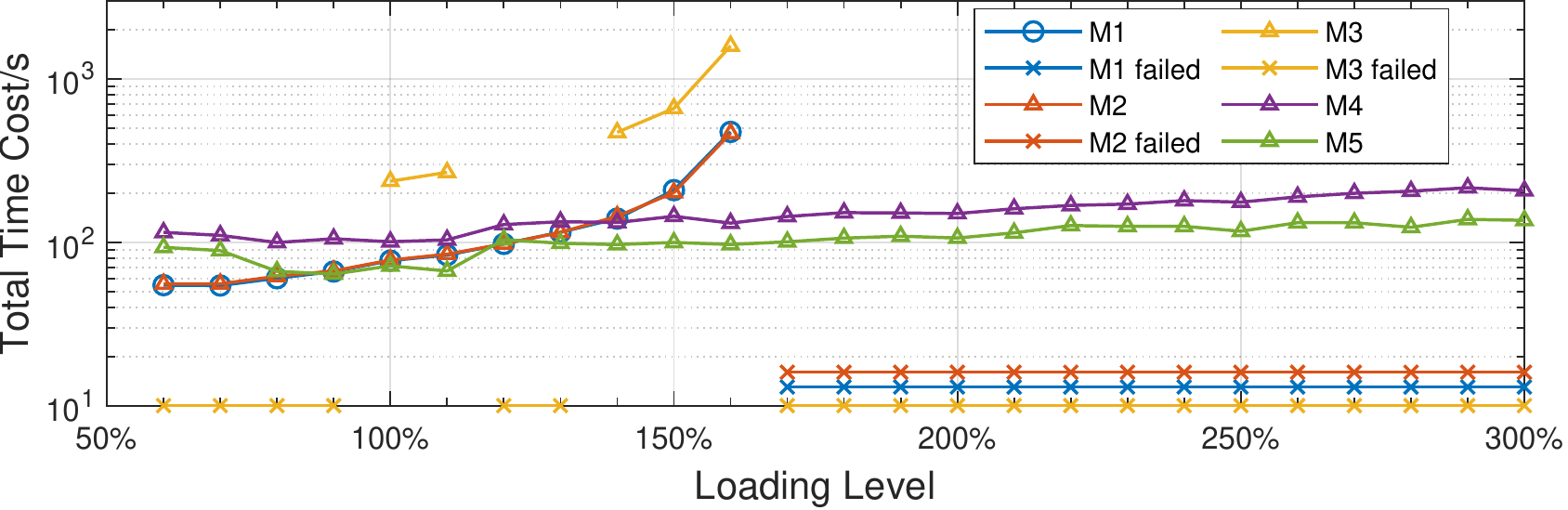}
    \caption{Total time costs under different loading level.}\label{paper DT figure robustness_total_time_costs}
\end{figure}
\begin{figure}[!h]
    \centering
    \includegraphics[width=3.4in]{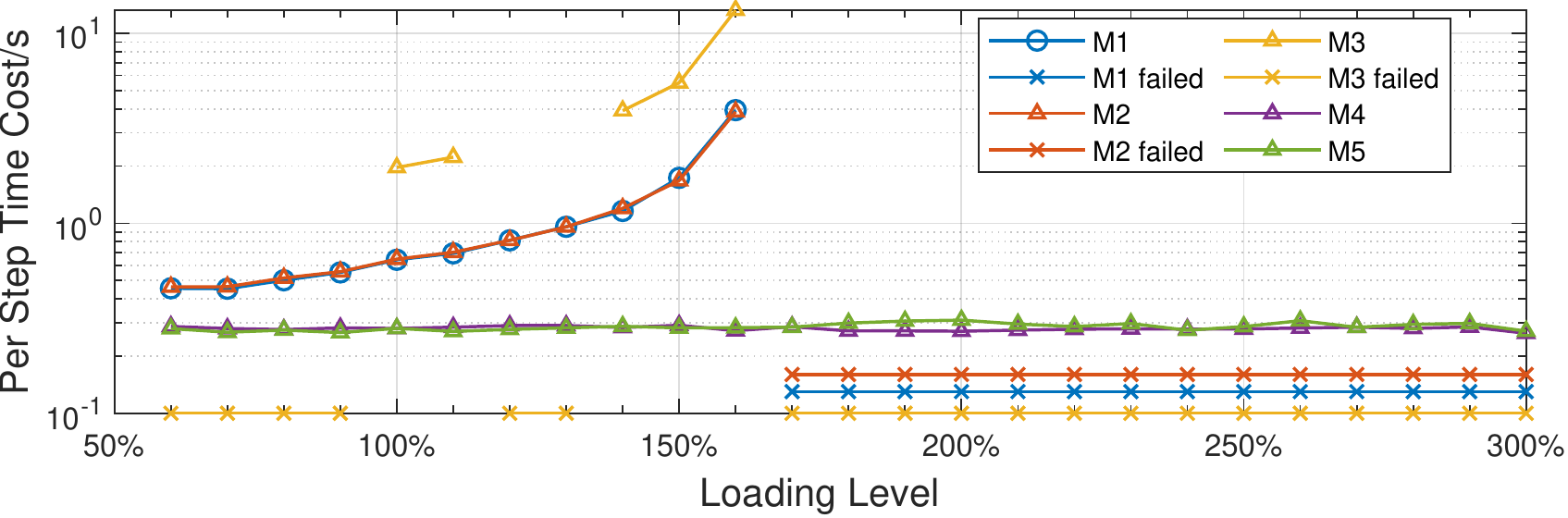}
    \caption{Per-step time costs under different loading level.}\label{paper DT figure robustness_per_time_costs}
\end{figure}
\begin{figure}[!h]
    \centering
    \includegraphics[width=3.4in]{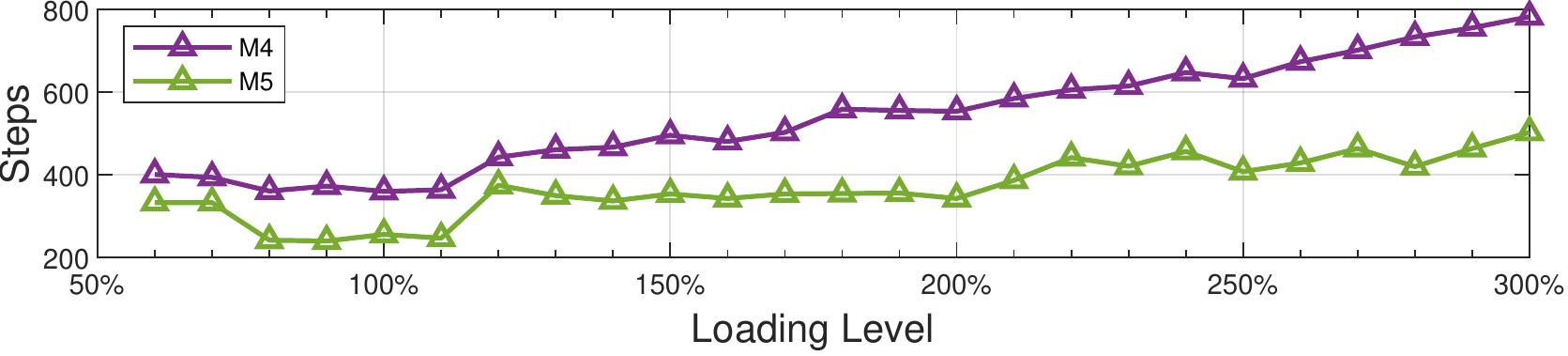}
    \caption{Number of steps under different loading level.}\label{paper DT figure robustness_steps}
\end{figure}
\begin{figure}[!h]
    \centering
    \includegraphics[width=3.4in]{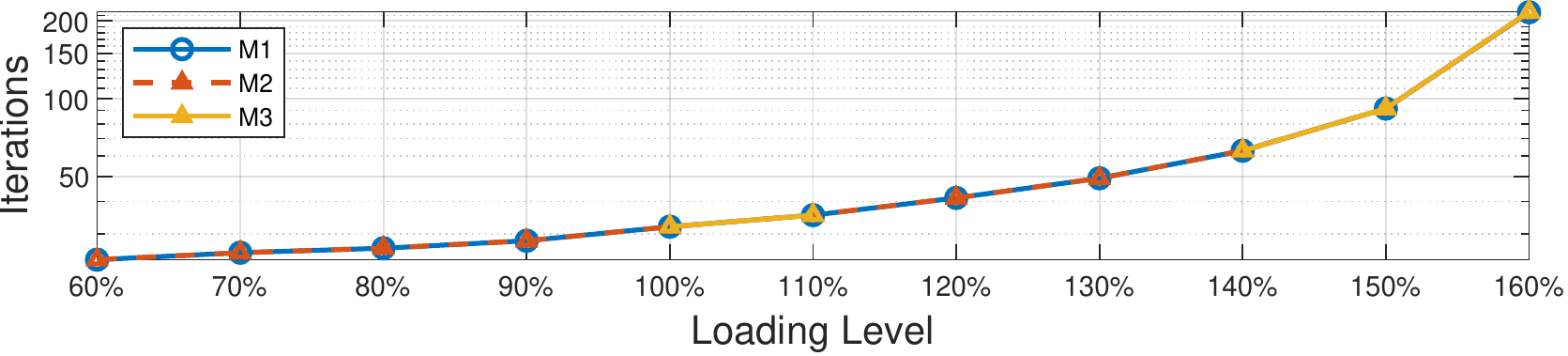}
    \caption{Average number of IES loop iterations under different loading level.}\label{paper DT figure robustness_iterations}
\end{figure}
\begin{figure}[!h]
    \centering
    \includegraphics[width=3.4in]{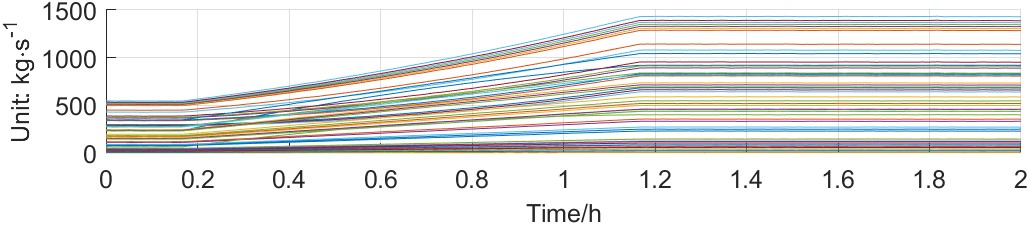}
    \caption{Mass flow rates by M4 when final loading level=300\%.}\label{paper DT figure robustness_mass_flow}
\end{figure}
\begin{figure}[!h]
    \centering
    \includegraphics[width=3.4in]{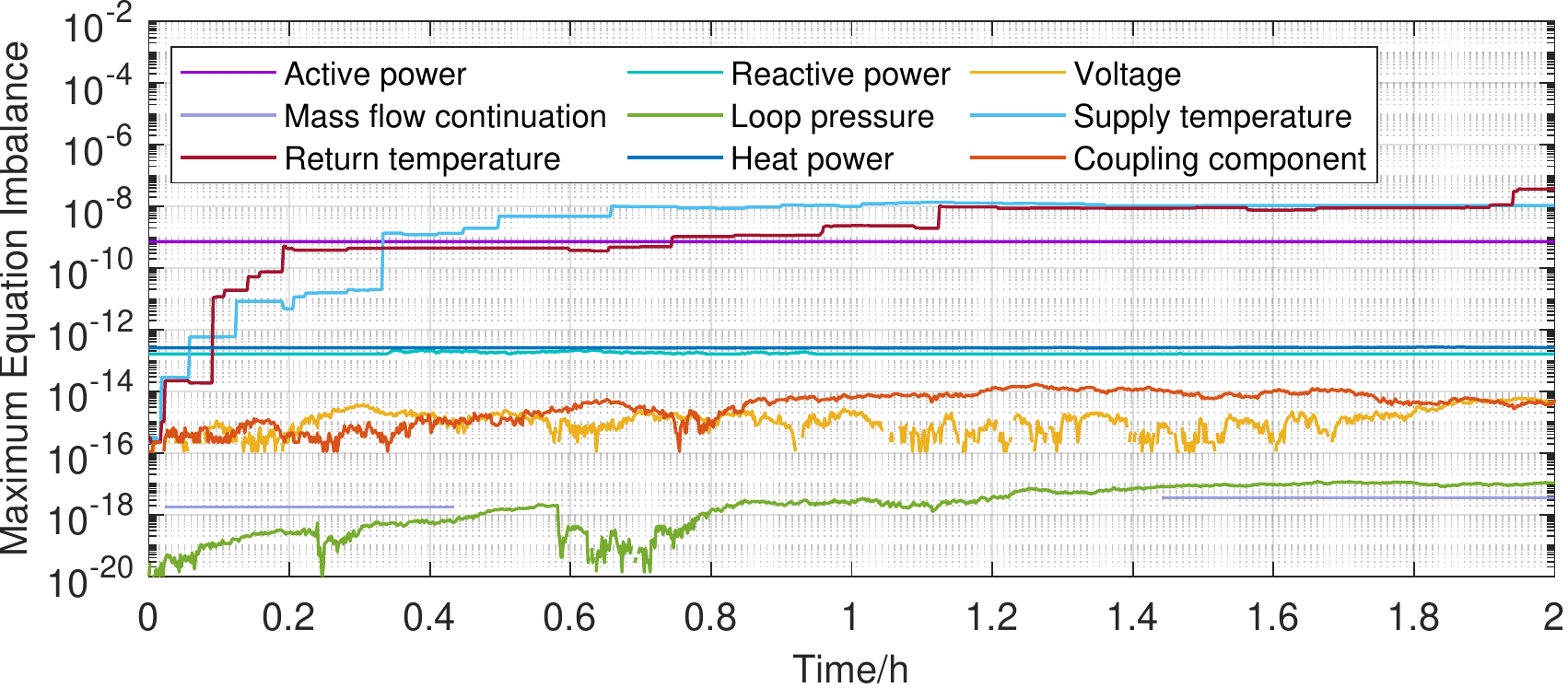}
    \caption{Maximum equation imbalance when final loading level=300\%.}\label{paper DT figure robustness_equation_balance}
\end{figure}
\subsubsection{Robustness}
In DHS, some heat load, for instance, the Guaranteed Saving Buildings (GSBs), spikes over ten times within one hour\cite{qinCombinedElectricHeat2021}. To test if the proposed method can produce convergent results and have robust time performance under this kind of severe disturbances, the following scenario is designed.
\begin{itemize}
    \item Scenario: $\mathcal{T}=$\SI{2}{\hour}. Loading level of all load nodes in the DHS changed from loading level 100\% at $t=$\SI{10}{\minute} to the target loading level, which ranges between 60\% and 300\% with step size 10\%, at $t=$\SI{70}{\minute}.
\end{itemize}
To simulate the real scenario, random noises with uniform distribution were added to the load curves and initial load values. The load curve when target loading level = 150\% is shown in Fig. \ref{paper DT figure robustness_phi}. M1-M5 were performed with previous settings.
\par 
The proposed method had consistent and robust time performance. As shown in Fig. \ref{paper DT figure robustness_per_time_costs}, the per-step time costs of M4 and M5 were nearly the same under all loading levels. As shown in Fig. \ref{paper DT figure robustness_total_time_costs} and \ref{paper DT figure robustness_steps}, the total time costs of M4 and M5 increased linearly as the target loading level grew since $\overline{\Delta t}$ diminished and the number of steps slightly increased. 
\par
In most cases, iteration solvers, M1-M3, produced divergent results. The total time costs of M1-M3 grew exponentially as the loading level grew. This is because the spiking loading level significantly increased the number of iterations and hence the per-step time costs. However, M1 and M2 were more efficient than the proposed method when the loading levels were smaller than 80\% for the number of iterations was tiny. M1-M3 had similar number of iterations in the convergent cases, but the per-step time cost of M3 was much bigger than that of M1 and M2. This is because M3 performed time-consuming index operations in each step. 
\par 
In the cases where M1-M3 and even the REF failed, we verified the correctness of the proposed method as follows. Firstly, Fig. \ref{paper DT figure robustness_mass_flow} is drawn to show that the results are reasonable, that is, they did not converge to meaningless solutions. Secondly, the results were verified by checking the imbalance of the nonlinear algebraic equations. We substituted the solutions into the original algebraic equations and it is shown in Fig. \ref{paper DT figure robustness_equation_balance} that the maximum equation imbalance is around 1e-8. Therefore, the proposed method was able to provide reliable results in this case, and hence demonstrated good robustness.
\begin{figure}
    \centering
    \begin{tikzpicture}
      \node[anchor=south west,inner sep=0] at (0,0) {\includegraphics[width=3.4in]{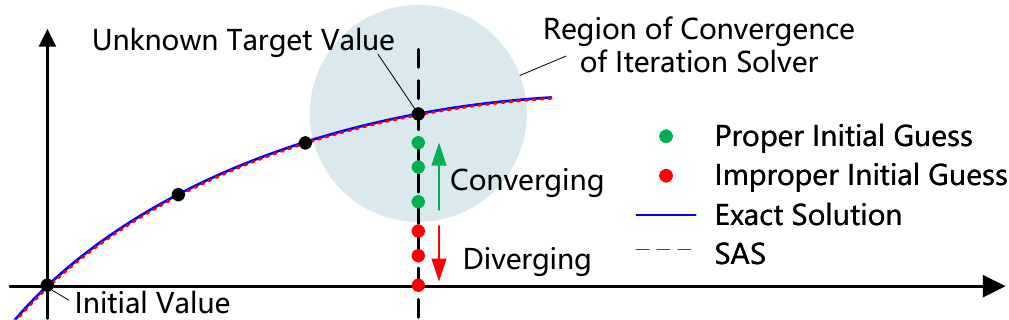}};
      \node (t) at (8.772,0.408) {$t$};
      \node (x) at (0.476,2.652) {$x(t)$};
      \node (t0) at (0.816,0.408) {$t_0$};
      \node (t') at (1.7,1.02) {$t'$};
      \node (t'') at (2.72,1.36) {$t''$};
      \node (t1) at (3.808,1.632) {$t_1$};
     \end{tikzpicture}\\
    \caption{Graphical explanation to the robustness of the proposed method.}\label{paper DT figure SAS VS NIM}
\end{figure}
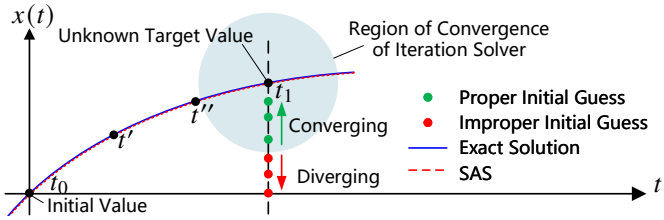
\par 
The possible explanations to the robustness superiority of the proposed method are as follows. 1) First, the proposed method solves DHS and EPS models together, getting rid of alternating errors. 2) Second, the adaptive time window control strategy controls the simulation errors within the prescribed error tolerance effectively. 3) Third, the proposed method does not rely on initial guesses while for iteration methods, initial guesses should be provided within the region of convergence. The region of convergence is an area probably centered by the unknown target value. However, the theory of iteration solver can not figure out the probable location of unknown target value, and the shape/area of the region of convergence. Following the common practice of solving differential equations, we assigned the results of the previous time window to the initial guesses when performing iteration methods. As shown in Fig. \ref{paper DT figure SAS VS NIM}, if the distances between these initial guesses and the target values are longer than the radius of the region of convergence, the iteration solver diverges. This accounts for the failures of iteration methods in the robustness tests where the variables changed violently and hence the target values are far away from the initial guesses. As shown in Fig. \ref{paper DT figure SAS VS NIM}, the proposed method starts the searching of the target value at $t=t_1$ from the initial value at $t=t_0$ with a definite trajectory. Though the approximation of target value may deviate greatly from the true values, we can use the adaptive time window control strategy to detect these errors, and perform simulation with smaller temporal step sizes again, for example, from $t=t'$ to $t=t''$. The above steps are conducted repeatedly until the value that the proposed method finds at $t=t_1$ is within the error tolerance. Therefore, the proposed method were more likely to ensure convergence in the preceding robustness tests.
\section{Conclusion}\label{paper DT section conclusion}
This paper proposes a DT-based non-iterative method to achieve efficient and robust time performance in HE-IES quasi-dynamic energy flow calculation. A semi-discrete TVD scheme is solved by DT to reduce dissipative and dispersive errors in thermal dynamics. An adaptive time window control strategy is designed to accelerate calculation and avoid non-convergence issues. 
\par
The proposed method has better accuracy performance compared with the FDM-based methods. But in cases where there are steep temperature changes, it cannot depict the temperature variations as accurately as NM. The proposed method has small and consistent per-step time costs regardless of loading levels, and thus, displays efficient and robust time performance. Whereas the iteration methods can only converge rapidly in lightly loaded cases. 
To choose parameter $\theta$ and order $K$ should also be careful.
In cases where there are steep temperature changes, we should set $\theta=1$ while in other cases, setting $\theta=2$ will be more efficient. Though small $K$ is suggested, we think that the choice of $K$ should be case-specific. Several tests should be performed ahead of simulation to avoid tiny average temporal step sizes.
The adaptive time window strategy is critical since the proposed method is conditionally stable.
\par 
The proposed method also applies to simplified scenarios where, for example, the DHSs are in quality regulation mode. By further considering the equipment models in HE-IESs, the proposed method is expected to perform efficient dynamic simulation, which will be our future research purpose. 
\appendices
\section{Step Sizes and Dissipative/Dispersive Errors}\label{paper DT appendix CFL Number, Dissipative and Dispersive Errors}
\renewcommand{\theequation}{\thesection.\arabic{equation}}
\setcounter{equation}{0}
We illustrate the relationships between step sizes and dissipative/dispersive errors using the following thermal-dynamics-like PDE
\[
    \frac{\partial \tau}{\partial t}+v\frac{\partial \tau}{\partial x}+0.9997(\tau-0.4)=0
\]
with step boundary condition. $v$ is the mass flow velocity.
\par
\begin{figure}[!h]
    \centering
    \subfloat[$\Delta x$=0.05, $\Delta t$=0.001, $v$=50, $R$=1]{\includegraphics[width=3in]{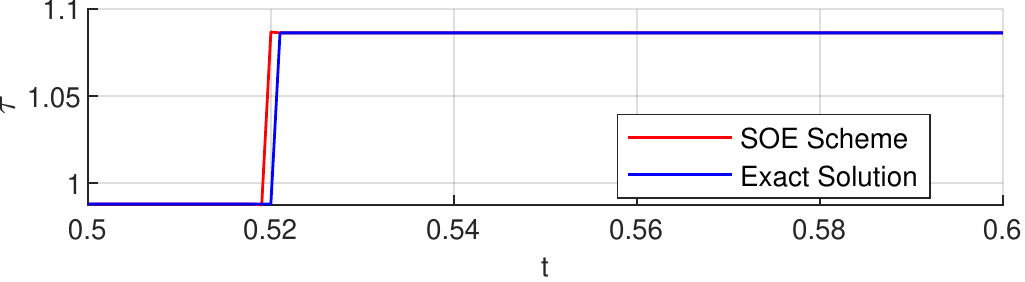}}\\
    \subfloat[$\Delta x$=0.05, $\Delta t$=0.001, $v$=55, $R$=1.1]{\includegraphics[width=3in]{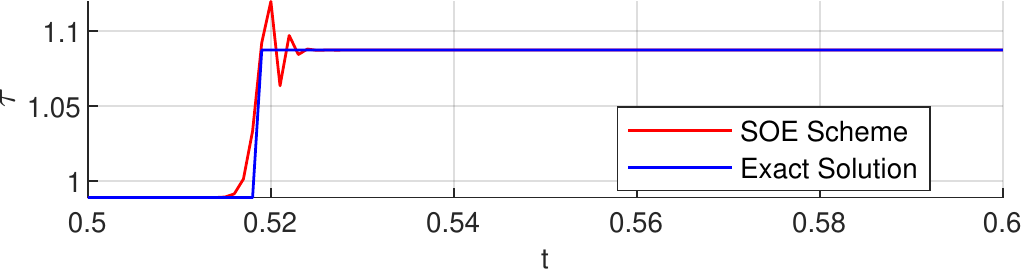}}
    \caption{Results by the SOE scheme.}\label{paper DT appendix A 1}
\end{figure}

 
For the SOE scheme \cite{Yao2021}
\begin{equation}\label{paper DT equation yao's scheme}
    \left\{
    \begin{aligned}
        &\frac{\partial \tau}{\partial t}=\frac{\tau_k^{n+1}-\tau_k^{n}+\tau_{k+1}^{n+1}-\tau_{k+1}^{n}}{2\Delta t}\\
        &\frac{\partial \tau}{\partial x}=\frac{\tau_{k+1}^{n+1}-\tau_k^{n+1}+\tau_{k+1}^{n}-\tau_{k}^{n}}{2\Delta x}
    \end{aligned}
    \right.,
\end{equation}
spatial step size $\Delta x$ and temporal step size $\Delta t$ satisfying $v\Delta t/\Delta x=1$ eliminate dissipative and dispersive errors in solutions effectively, which is shown in Fig. \ref{paper DT appendix A 1}(a). $R=v\Delta t/\Delta x$ is called Courant-Friedrichs-Lewy (CFL) number in literature and serves as an important stability indicator of difference schemes\cite{Thomas1995}. However, if we increase $v$ slightly, then the dispersive errors, which are the fake oscillations in Fig. \ref{paper DT appendix A 1}(b), become obvious. Therefore, SOE scheme is able to obtain accurate results when mass flow velocities are fixed, but it is difficult for the scheme to avoid dispersive errors when mass flow velocities are variable. 
\par
\begin{figure}[!h]
    \centering
    \subfloat[$\Delta x$=0.05, $\Delta t$=0.001, $v$=50, $R$=1]{\includegraphics[width=3in]{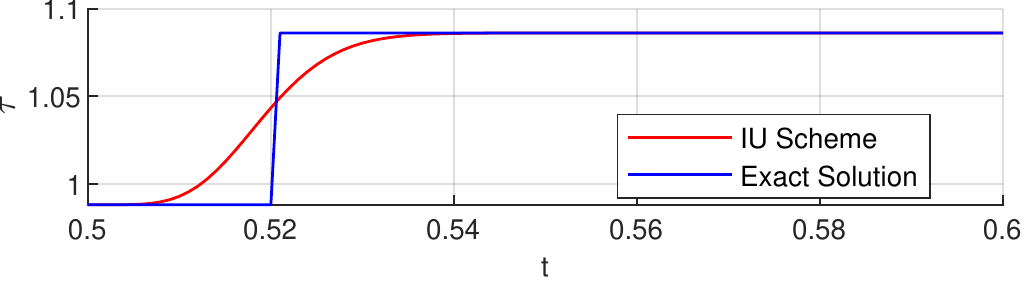}}\\
    \subfloat[$\Delta x$=0.001, $\Delta t$=0.001, $v$=50, $R$=50]{\includegraphics[width=3in]{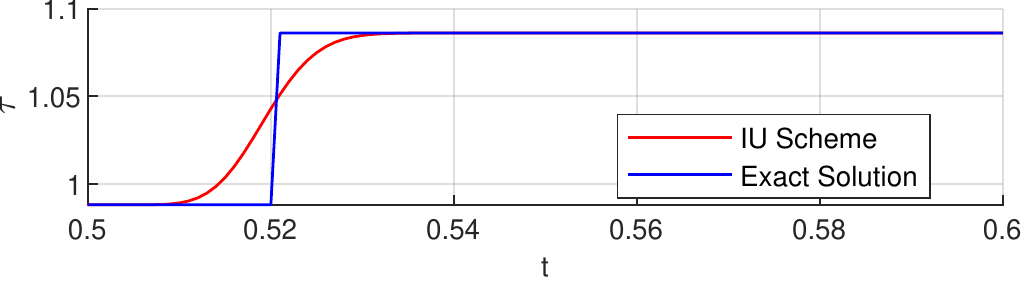}}
    \caption{Results by the IU scheme.}\label{paper DT appendix A 2}
\end{figure}

For the IU scheme \cite{Wangyaran2017}
\begin{equation}\label{paper DT equation implicit upwind scheme}
    \left\{
    \begin{aligned}
        &\frac{\partial \tau}{\partial t}=\frac{\tau_{k+1}^{n+1}-\tau_{k+1}^{n}}{\Delta t}\\
        &\frac{\partial \tau}{\partial x}=\frac{\tau_{k+1}^{n+1}-\tau_k^{n+1}}{\Delta x}
    \end{aligned}
    \right.,
\end{equation}
it can be observed from Fig. \ref{paper DT appendix A 2} that neither putting $R=1$ nor decreasing $\Delta x$ can eliminate the decaying of high-frequency components, which are the manifestation of dissipative errors in solutions. 
Therefore, solutions by the IU scheme are always accompanied by the dissipative errors.
\section{Notation Explanation}
\label{paper DT appendix Notation Explanation}
\begin{figure}[h]
    \centering
    \begin{tikzpicture}
        \node[anchor=south west,inner sep=0] at (0,0) {\includegraphics[width=2.8in]{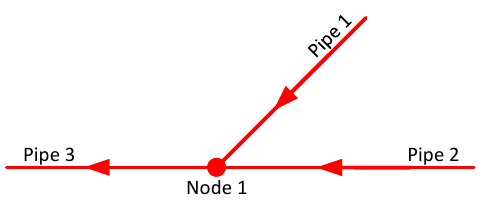}};
        \node (T1) at (2.8,0.8) {$\taus_1$};
        \node (Tout2) at (4.1,0.8) {$\tauouts_2$};
        \node (Tout1) at (3.4,1.1) {$\tauouts_1$};
        \node (m2) at (5,0.8) {$\m_2$};
        \node (m1) at (4,1.9) {$\m_1$};
        \node (m3) at (1.6,0.8) {$\m_3$};
    \end{tikzpicture}\\
    \caption{Temperature mixture case in the supply network.}\label{paper DT figure notation of node temperature equation1}
\end{figure}
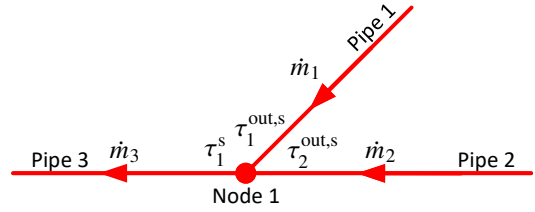

Notation of \eqref{paper DT AE node temperature equations s mixture} and \eqref{paper DT AE node temperature equations r mixture} can be explained using the supply network example in Fig. \ref{paper DT figure notation of node temperature equation1} where we assume Node 1 to be an intermediate node.
\par
$\taus_1$ can be calculated by node temperature mixture equation \cite{Qin2019} as 
\[\taus_1=\tauouts_1\frac{\m_1}{\m_1+\m_2}+\tauouts_2\frac{\m_2}{\m_1+\m_2}.\]
Moving $\m_1+\m_2$ in denominator to the left hand side, we have 
\[\taus_1(\m_1+\m_2)=\tauouts_1\m_1+\tauouts_2\m_2.\]
Adding $\m_3$ and $\tauouts_3$ to the equation, and rewriting it as matrix-vector form, we have 
\[\taus_1
\begin{bmatrix}
    1&1&0
\end{bmatrix}
\begin{bmatrix}
    \m_1\\ 
    \m_2\\ 
    \m_3\\
\end{bmatrix}
=
\begin{bmatrix}
    1&1&0
\end{bmatrix}
\begin{bmatrix}
    \tauouts_1&&\\ 
    &\tauouts_2&\\ 
    &&\tauouts_3\\
\end{bmatrix}
\begin{bmatrix}
    \m_1\\ 
    \m_2\\ 
    \m_3\\
\end{bmatrix}.
\]
Because
\[\begin{bmatrix}
    1&1&0
\end{bmatrix}=\max(V_\text{1},\begin{bmatrix}
    0&0&0
\end{bmatrix})=V_\text{1}^\text{+}\]
where $V_\text{1}=\begin{bmatrix}
     1&1&-1
\end{bmatrix}$ denotes the row of node incidence matrix related to the intermediate node---Node 1, we have 
\[
    \taus_1 V_\text{1}^\text{+}\bm=V_\text{1}^\text{+}\md(\btau^\text{out,s})\bm.
\]
Node supply/return temperature equations \eqref{paper DT AE node temperature equations s mixture} and \eqref{paper DT AE node temperature equations r mixture} for other types of nodes can be obtained in a similar way.
\bibliographystyle{IEEEtran}
\bibliography{../../../latex/IEEEabrv,../../../latex/bibliography}
\begin{IEEEbiography}[{\includegraphics[width=1in,height=1.25in,clip,keepaspectratio]{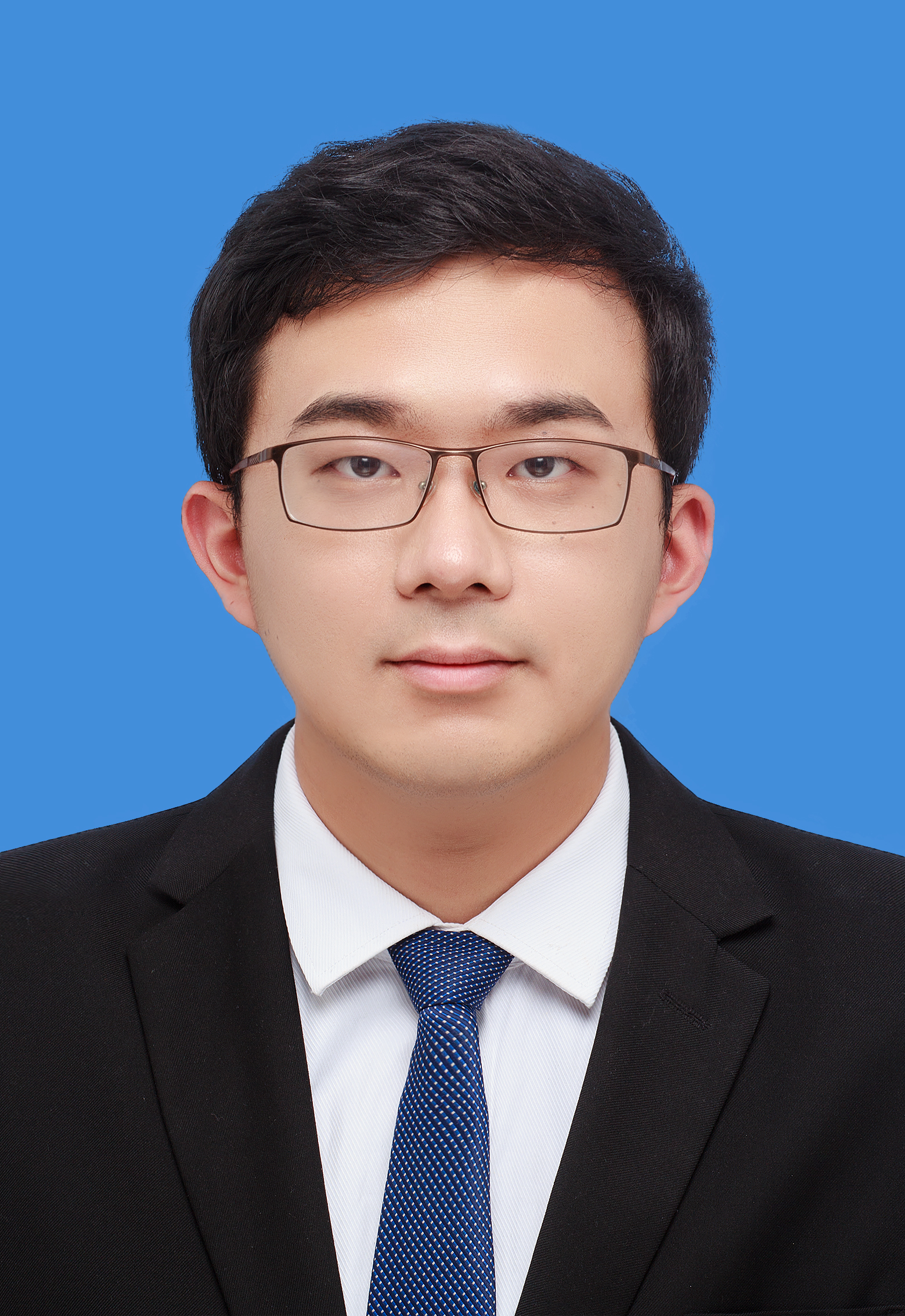}}]{Ruizhi Yu}(S'20) received the B.S.E.E. degree in 2020 from the Chien-Shiung Wu honors college of Southeast University, Nanjing, China, where he is currently working toward the Ph.D.E.E. degree. His research interests include simulation and scripting of integrated energy systems.
\end{IEEEbiography}
\begin{IEEEbiography}[{\includegraphics[width=1in,height=1.25in,clip,keepaspectratio]{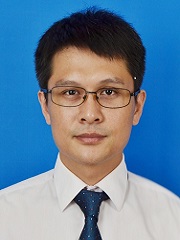}}]{Wei Gu}(M'06-SM'16) received his B.S. and Ph.D. degrees in Electrical Engineering from Southeast University, China, in 2001 and 2006, respectively. From 2009 to 2010, he was a Visiting Scholar in the Department of Electrical Engineering, Arizona State University. He is now a professor in the School of Electrical Engineering, Southeast University the director of the Institute of Distributed Generations and Active Distribution Networks. His research interests include distributed generations and microgrids, integrated energy systems. Dr. Gu is an Editor for the IEEE Transactions on Power Systems, the IET Energy Systems Integration and the Automation of Electric Power Systems (China).
\end{IEEEbiography}
\begin{IEEEbiography}[{\includegraphics[width=1in,height=1.25in,clip,keepaspectratio]{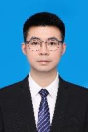}}]{Suhan Zhang}(S'17-M'21) received his B.S. degree in electrical engineering from North China Electric Power University, China, in 2018. He is currently working toward the Ph.D. degree from the School of Electrical Engineering, Southeast University, Nanjing, Jiangsu, China. His research interests include modeling, simulation and operation of integrated energy systems.
\end{IEEEbiography}  
\begin{IEEEbiography}[{\includegraphics[width=1in,height=1.25in,clip,keepaspectratio]{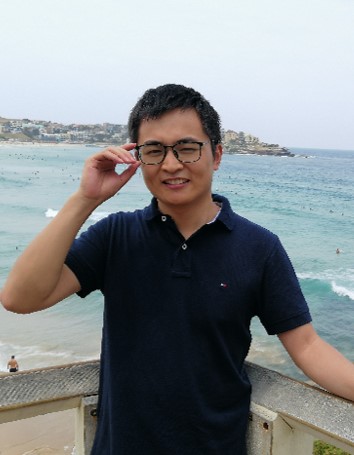}}]{Shuai Lu}(S'17-M'21) received his B.S. degree in Smart Grid Information Engineering from Nanjing University of Science and Technology, Nanjing, China, in 2016 and his Ph.D. degree in Electrical Engineering from Southeast University, Nanjing, China, in 2021. From 2018 to 2019, he was a visiting scholar at the University of New South Wales, Sydney, Australia. He is currently a Lecturer at the School of Electrical Engineering, Southeast University. He was selected as an Outstanding Reviewer for IEEE Transactions on Power Systems in 2020. His research interests include multi-energy systems, operations research, and data-driven techniques in power systems.  
\end{IEEEbiography}  
\begin{IEEEbiography}[{\includegraphics[width=1in,height=1.25in,clip,keepaspectratio]{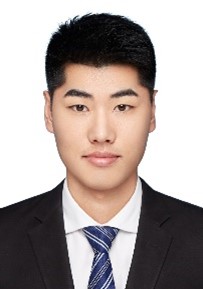}}]{Shixing Ding}(S'20) received his M.S. degree in Power Engineering from North China Electric Power University, Beijing, China, in 2019. He is currently pursuing a Ph.D. degree in Cyber Science and Engineering at Southeast University, Nanjing, China. His research interests include modeling, simulation, and optimization of integrated energy systems; CPS security for energy systems.    
\end{IEEEbiography}
\end{document}